\documentclass[11pt,a4paper]{article}

\usepackage[title]{appendix}

\usepackage{jheppub}
\usepackage{graphicx} % Required for inserting images
\usepackage{color}
\usepackage{amsmath}
\usepackage{listings}
\usepackage{exercise}
\usepackage{amssymb}
\usepackage{cases}
\usepackage{graphicx}
\usepackage{hyperref}

\hypersetup{
	colorlinks=true,
	linkcolor=blue,
	filecolor=blue,      
	urlcolor=blue,
	citecolor=blue
}

\usepackage{fancyhdr}
\usepackage{mdframed}
\mdfdefinestyle{exampledefault}{
	outerlinewidth=5pt,innerlinewidth=0pt,
	outerlinecolor=red,roundcorner=5pt
}
\usepackage{empheq}
\usepackage{xcolor}
\usepackage[compat=1.0.0]{tikz-feynman}
\usepackage{tcolorbox}
\usepackage{cancel}
\usepackage{subcaption}

\usepackage[compat=1.0.0]{tikz-feynman}
\usepackage{tcolorbox}
\usepackage{soul}

\usepackage{jheppub}
\usepackage{graphicx}
\usepackage{amsmath,amssymb,amsthm}
\usepackage{physics}
\usepackage{enumitem}
\usepackage{caption,subcaption}
\usepackage{hyperref}
\usepackage{cleveref}
\usepackage{xcolor}
\usepackage{empheq}
\graphicspath{{./img/}}

\def\beq{\begin{equation}}
\def\eeq{\end{equation}}
\def\bea{\begin{eqnarray}}
\def\eea{\end{eqnarray}}
\def\bitem{\begin{itemize}}
\def\eitem{\end{itemize}}
\newcommand{\newc}{\newcommand}
\newc{\gsim}{\lower.7ex\hbox{$\;\stackrel{\textstyle>}{\sim}\;$}}
\newc{\lsim}{\lower.7ex\hbox{$\;\stackrel{\textstyle<}{\sim}\;$}}

\begin{document}

\title{A more effective QCD string at colliders: Decay of excited strings and the worldsheet axion}

\author{Ethan Carragher,}
\author{John March-Russell}

\affiliation{Rudolf Peierls Centre for Theoretical Physics, University of Oxford, OX1 3PU, UK}

\emailAdd{ethan.carragher@physics.ox.ac.uk}
\emailAdd{john.march-russell@physics.ox.ac.uk}

\abstract{
The confining flux tube of $(3+1)$d QCD is described by an effective string theory with $(1+1)$d worldsheet action that extends the Nambu-Goto form by the addition of a massive pseudoscalar worldsheet ``axion".  As argued in companion papers concerning the modified phenomenology of the Lund string model at colliders, QCD flux tubes produced by high-energy collisions are likely to involve excitation of both worldsheet Nambu-Goldstone and axion modes, although the standard Lund model assumes a constant tension ground-state string.   Here we detail the path-integral computation of the modified Schwinger-like process of string breaking via nucleation of quark-antiquark pairs in the presence of excitations above the string ground state. We find that the worldsheet axion leads to the dominant change in the string breaking process, the axion excitations producing, among other effects, a varying effective tension of the string, which can exponentially enhance or suppress the string breaking rate depending on the local phase of the excitation.  Our computation employs a version of the Schwinger-Keldysh complex time contour method with initial state data specified by a density matrix. In an excited background the Euclidean saddle point is generically complex, but its continuation gives real initial data for post-decay evolution.  Our results are of relevance for hadronisation models with excited QCD strings.}

\date{\today}

\maketitle

\section{Introduction}
\label{sec:intro}

The Lund string model provides the standard description of hadronisation in modern collider event generators such as \textsc{Pythia} and underlies much of the quantitative interpretation of QCD final states~\cite{Andersson:1983ia,Sjostrand:2014zea}. In this framework, confinement is modelled by a string-like flux tube connecting coloured partons, and hadron production occurs through repeated string breaking via quark-antiquark pair creation. The corresponding production rate per unit length is conventionally parameterised by the Schwinger-like expression
\begin{align}
\frac{d\Gamma}{d\ell} \propto \exp\!\left(-\frac{\pi m_q^2}{\kappa}\right),
\label{Eq:Lund}
\end{align}
where $\kappa$ denotes the string tension and $m_{q}$ an effective endpoint mass parameter. Despite its phenomenological success, this ansatz is not derived within a systematically improvable framework. In particular, it implicitly assumes that the string remains in its ground state, neglecting the possible impact of worldsheet excitations on the string-breaking process.

Such an assumption is difficult to justify for flux tubes relevant for hadronisation, as they are produced in high-energy collisions and need not form in their lowest-energy state. It is therefore important to quantify how excited worldsheet backgrounds modify the string breaking behaviour.

A controlled framework for addressing this question is provided by the effective theory of the confining string. Long flux tubes are described at low energies by a $(1+1)$-dimensional worldsheet theory whose predictions successfully reproduce universal features of flux-tube spectra such as the L{\"u}scher correction~\cite{Arvis:1983fp,Luscher:1980fr,LUSCHER1981317,Luscher:2004ib}, as supported by extensive lattice studies~\cite{Lucini:2001nv,Athenodorou:2011rx,Athenodorou:2010cs}. Within this framework, string breaking admits a controlled semi-classical treatment closely analogous to false-vacuum decay in quantum field theory~\cite{Coleman:1977py,Callan:1977pt}.\footnote{An even closer analogy is the bubble-of-nothing decay process of space-time~\cite{Witten:1981gj,Dine:2004uw,GarciaEtxebarria:2020xsr,Draper:2021ujg}.} The process can be formulated as the decay of a metastable string worldsheet, proceeding through the nucleation of a quark-antiquark pair~\cite{Monin:2008mp}. In Euclidean signature this corresponds to a ``bounce" configuration in which a circular hole is excised from the string worldsheet, and the decay rate is determined by the associated bounce action. This reproduces the Lund result Eq.~\eqref{Eq:Lund} for a string in its ground state, while simultaneously providing a systematic starting point for computing corrections due to worldsheet excitations.

By far the best understood degrees of freedom on the worldsheet are the two massless Nambu–Goldstone boson (NGB) modes associated with transverse string fluctuations. Their effects on string breaking have been investigated in a variety of settings. Thermal NGB backgrounds at low temperatures modify only the fluctuation determinant around the bounce and leave the exponential suppression factor unchanged~\cite{Monin:2008mp,Monin:2008uj}, while they do alter the exponent at sufficiently high temperatures~\cite{Garriga:1994ut}. Individual NGB quanta may likewise catalyse decay through scattering processes, affecting the exponent only at high energies~\cite{Monin:2009ch}. Importantly, NGB-induced corrections are parametrically suppressed at leading order in the derivative expansion of the effective string theory. First derivatives of the NGB fields correspond only to local boosts and transverse rotations of the worldsheet, and therefore cannot modify the intrinsic decay rate because of the underlying $(3+1)$-dimensional Poincar{\'e} invariance. Corrections to the decay rate arise only from higher-derivative operators and are consequently suppressed by powers of the string tension.

Recent precision lattice studies have revealed evidence for a qualitatively different excitation on the QCD string worldsheet: a massive pseudoscalar state known as the \emph{worldsheet axion}~\cite{Dubovsky:2013gi,Athenodorou:2024loq,Sharifian:2025fyl}. Current evidence suggests this mode to be a robust feature of confining $SU(N_c)$ strings in four spacetime dimensions, both with and without quarks. Its worldsheet action contains a coupling to a topological density sensitive to self-intersections~\cite{Dubovsky:2012sh,Dubovsky:2013gi,Dubovsky:2015zey}. For long, weakly curved strings, however, this interaction is subleading, allowing the axion to be treated approximately as a free massive scalar propagating on the worldsheet. Unlike the NGB modes, the axion is not constrained by the symmetry argument described above, raising the possibility of parametrically larger effects on the string-breaking rate.

Determining these effects requires going beyond the standard Euclidean bounce formalism of false vacuum decay. The usual derivation assumes evolution over infinite Euclidean time, thereby projecting onto the ground state and eliminating all information about initial excitations. To retain the effect of a specified excited state, the problem must instead be formulated using an in-in path integral. This leads naturally to a Schwinger–Keldysh contour in complex time, with the initial state encoded by a density matrix. The string-breaking rate is then obtained from a saddle-point evaluation of a doubled path integral, generalising the standard bounce construction.

In this work we apply this formalism to compute the leading modification of the string-breaking rate induced by a semi-classical worldsheet axion background. We show that the decay rate depends on an effective spacetime-dependent tension determined by the local value of the worldsheet axion field and its derivatives, at least in the long-wavelength regime. Depending on the phase of the excitation, the decay rate can be exponentially enhanced or suppressed. These effects are generically larger than those arising from NGB backgrounds and may have significant phenomenological consequences for hadronisation, including enhanced production of heavier quark flavours. While our primary motivation is the phenomenology of excited QCD strings produced at colliders, the formalism developed here applies more broadly to tunneling processes in excited backgrounds.

The possibility that excited strings may exhibit modified fragmentation dynamics has been considered previously. Earlier studies have investigated the phenomenological consequences of strings with time-dependent effective tensions~\cite{Hunt-Smith:2020lul}, while other approaches have invoked interactions between multiple strings, colour reconnection, or collective flux-tube effects~\cite{ANDERSSON199182,Bierlich:2014xba,Christiansen:2015yqa,Duncan:2019poz,Altmann:2025afh}. The present work differs in that it considers a single isolated string carrying a specified worldsheet excitation and derives the resulting modification of the decay rate directly within the underlying worldsheet effective theory, thereby allowing for systematic improvement.

Finally, we note that confining Yang-Mills theories support more than one type of locally stable flux tube, characterised for example by their transformation properties under the centre of the gauge group. Such strings generically possess different tensions and may exhibit additional dynamical features~\cite{Lucini:2001nv,Lucini:2004my}. Confining strings can also support interesting structures such as baryon junctions, which can be incorporated in the effective theory~\cite{Komargodski:2024swh} and give rise to characteristic breaking patterns~\cite{Altmann:2024odn}. In the present work we restrict attention to the fundamental flux tube of $SU(N_c)$ and leave the investigation of other string sectors and worldsheet defects to future studies.

The paper is organised as follows. In \Cref{Sec:Action} we review the effective worldsheet theory, including the axion mode. In \Cref{Sec:GroundState} we review the standard bounce calculation for ground-state string breaking. In \Cref{sec:pairproductionraterigorous} we develop the Schwinger–Keldysh formalism appropriate for excited worldsheet backgrounds. We apply this to compute the resulting decay rate in an axion background for the simple case of a massless axion in \Cref{sec:simplecases}, and the more general case $0 < m_{a} R \ll 1$ in \Cref{sec:generalsolution}. We conclude in \Cref{sec:Conclusions}.

\section{Effective theory of the QCD string}
\label{Sec:Action}

We begin by briefly reviewing the (1+1)-dimensional worldsheet description of the open string with massive ``quark/antiquark" endpoints.\footnote{For simplicity we take the endpoints to be spinless, since this does not affect the later calculation of the string breaking rate; it is possible to incorporate endpoint spin using the methods outlined in \cite{Cuomo:2024gek}.} The effective theory of confining strings has been developed extensively over the past two decades~\cite{Aharony:2009gg,Aharony:2010cx,Aharony:2010db,Aharony:2011gb,Aharony:2011ga,Dubovsky:2012sh,Dubovsky:2014fma,Dubovsky:2015zey,Dubovsky:2016cog,Cuomo:2024gek}; for detailed reviews see~\cite{Aharony:2013ipa,Brandt:2016xsp}.
The string worldsheet $\mathcal{W}$ is parameterised by coordinates $\xi^{\alpha}$, $\alpha \in \{0,1\}$, and embedded into $(3+1)$-dimensional Minkowski spacetime through coordinate fields $X^{\mu}(\xi^{\alpha})$, with metric $\eta_{\mu\nu} = \text{diag}(1,-1,-1,-1)$. Worldsheet reparameterisation invariance implies that the action must be constructed from geometric quantities, most importantly the induced metric
\begin{align}
    h_{\alpha \beta} = \eta_{\mu \nu} \partial_{\alpha} X^{\mu} \partial_{\beta} X^{\nu} \equiv \partial_{\alpha} X \cdot \partial_{\beta} X,
\end{align}
which defines the worldsheet line element
\begin{align}
    ds^{2} = h_{\alpha\beta}\, d\xi^{\alpha} d\xi^{\beta}.
\end{align}
At leading order in the derivative expansion, the effective worldsheet action consists of the Nambu-Goto bulk term together with a massive endpoint contribution,
\begin{align}
    S_{0} = - \kappa \int_{\mathcal{W}} d^{2}\xi\, \sqrt{- h} - m_{q} \int_{\partial \mathcal{W}} ds,
    \label{eqn:NGaction}
\end{align}
where $\kappa$ is the string tension and $m_{q}$ is an effective endpoint mass parameter. Physically, $m_{q}$ should be interpreted as the mass of a dressed endpoint quark, including the gluonic fields located near the end of the flux tube, rather than either the bare electroweak quark mass or the constituent quark mass of hadronic phenomenology. The values used in the \textsc{Pythia} implementation of the Lund string model are discussed in our companion paper.

Here we study the string breaking process as a function of $\kappa,\, m_{q}$, and the parameters of the worldsheet axion introduced below. Throughout we assume they lie in a regime in which the semi-classical description is reliable. We require in particular $m_{q} / \sqrt{\kappa} \gg 1$, which allows relatively long-lived metastable string states. This condition further ensures the endpoints can be treated as pointlike within the effective theory.

The action Eq.~\eqref{eqn:NGaction} is invariant under global target-space Poincar{\'e} transformations acting on the embedding fields $X^\mu$. For a long straight string, this symmetry is spontaneously broken according to $IO(3,1) \rightarrow IO(1,1)\times O(2)$, corresponding to the preservation of Poincar{\'e} transformations along the string together with rotations in the transverse plane. The resulting low-energy worldsheet theory contains two independent massless Nambu-Goldstone boson (NGB) modes associated with transverse fluctuations of the string.\footnote{Although several spacetime generators are spontaneously broken, not all give rise to independent Nambu-Goldstone fields. After accounting for the relations among the broken spacetime symmetries, only two physical massless modes remain.} These degrees of freedom are most transparently exhibited in ``static" gauge,\footnote{This gauge choice is valid provided the string remains close to a straight configuration, without back-tracking or self-intersections, which is sufficient for our purposes.}
\begin{align}
    X^\mu(t,x) = \left(t,\, x,\, \frac{\tilde X^{2}(t,x)}{\sqrt{\kappa}},\, \frac{\tilde X^{3}(t,x)}{\sqrt{\kappa}} \right),
    \label{eqn:physicalgauge}
\end{align}
which may be viewed as a unitary gauge for the worldsheet reparameterisations. Expanding the bulk NG action in derivatives of the transverse fluctuations yields
\begin{equation}
    S_{NG} = \int dt dx\, \left[-\kappa + \frac{1}{2} \partial_{t} \tilde X^{i} \partial_{t} \tilde X^{i} - \frac{1}{2} \partial_{x} \tilde X^{i} \partial_{x}\tilde X^{i} + \mathcal{O}\left(\frac{(\partial \tilde X)^{4}}{\kappa} \right) \right],
\label{eqn:action_NG_expanded}
\end{equation}
where a sum over $i = 2,3$ is implied. The fields $\tilde X^{i}$ are therefore the massless Nambu-Goldstone degrees of freedom.

In addition to these massless modes, lattice studies of confining strings in $D=4$ spacetime dimensions (though not in $D=3$) provide strong evidence for an additional massive pseudoscalar excitation on the worldsheet, commonly referred to as the \emph{worldsheet axion}~\cite{Dubovsky:2013gi}. This state is observed for open and closed confining strings, including those of $SU(3)$ Yang-Mills theory~\cite{Athenodorou:2010cs,Dubovsky:2013gi,Athenodorou:2021vkw,Sharifian:2023idc,Athenodorou:2024loq,Sharifian:2024nco,Athenodorou:2025mbu}. The axion mass is approximately $m_{a} \approx 1.85 \sqrt{\kappa}$. Current precision lattice data~\cite{Athenodorou:2024loq}, as well as fits to the spectra of high-spin mesons~\cite{Cuomo:2024gek}, do not show evidence for additional light worldsheet degrees of freedom.

The worldsheet axion is odd under both longitudinal and transverse spacetime parity transformations, and therefore transforms as a pseudoscalar on the worldsheet.\footnote{The relevant discrete symmetries arise from the reflection components of the unbroken subgroups $IO(1,1)$ and $O(2)$, corresponding to longitudinal and transverse parities $P_\perp$ and $P_\parallel$, respectively. The symmetry $P_\parallel$ follows from the unbroken $CP$ symmetry of $SU(N_c)$ gauge theory (assuming vanishing $\theta$-angle), since it exchanges the quark and antiquark endpoints.} The leading-order addition to the effective action involving the axion field $a$ is~\cite{Dubovsky:2015zey}
\begin{align}
    S_{\text{a}} = \frac{1}{2}\int dt dx\, \sqrt{-h} \,\biggl(h^{\alpha \beta} \partial_{\alpha} a \partial_{\beta} a - m_{a}^{2} a^{2}+\ldots\biggr) + S_{\mathrm{int}}\, ,
\label{Eq:effective_action}
\end{align}
where the interaction term is
\begin{align}
    S_{\mathrm{int}} = \frac{Q}{8 \pi} \int d^{2}\xi\, a\, \varepsilon^{\alpha \beta} \varepsilon_{ij} K^{i}_{\alpha \gamma} K^{j \gamma}_{\beta}.
\end{align}
Here $K^{i}_{\alpha \gamma} = \nabla_{\alpha} \partial_{\gamma} X^{i}$ is the extrinsic curvature of the worldsheet, and $Q$ is a dimensionless coupling. Fits to lattice data suggest $Q \approx 9.6$, remarkably close to the value associated with an integrable massless-axion worldsheet theory~\cite{Dubovsky:2013gi}. Interestingly, the interaction term is proportional to the self-intersection density of the worldsheet which integrates to a topological constant for constant $a(t,x)$.\footnote{This observation suggests that the worldsheet axion may be a compact field, though we will not pursue the possible consequences of this periodicity here.} However, since the straight-string configurations relevant to this work have vanishing self-intersection number, we neglect $S_{\rm int}$ in what follows. Also note that the axion does not interact with the endpoints at leading order.

Equations \eqref{eqn:NGaction} and \eqref{Eq:effective_action} should be understood as the leading terms in the low-energy effective field theory of the confining string. The expansion is organised in derivatives of $X^\mu$ and in derivatives and powers of $a(t,x)$, and is expected to break down above a physical UV cutoff $\Lambda \sim \sqrt{2\pi \kappa}$. Above this scale, the effective string description must be matched onto the underlying gauge theory of quarks and gluons. Furthermore, quantisation\footnote{Contrary to a common misconception, the NG action can be consistently quantised outside the critical bosonic-string dimension, $D=26$, when considered as the \emph{leading term of an effective field theory}~\cite{Dubovsky:2012sh}. The way this is achieved depends on the gauge choice for the worldsheet
reparameterization symmetry. In static gauge, unitarity is manifest but spacetime Poincar{\'e} symmetry is realised nonlinearly; verifying closure of the symmetry algebra requires an appropriate regulator, such as dimensional regularisation on the (1+1)d worldsheet, together with evanescent operators that identically vanish in $d = 2$ but not in $d= 2-2\epsilon$. In conformal gauge, by contrast, Poincar{\'e} symmetry is manifest while unitarity is less transparent, and consistency requires the inclusion of the Polchinski-Strominger term \cite{Polchinski:1991ax}.} predicts the appearance of a tachyonic instability once the string length becomes shorter than $\sim 1/\sqrt{\kappa}$, so strictly speaking the effective description is controlled only in the long-string regime. Nevertheless, the NG-plus-axion theory successfully reproduces a wide range of detailed lattice results even for strings with lengths not far above $1/\sqrt{\kappa}$.

As explained in the introduction, the dominant correction to the string-breaking rate does not arise from the NGBs, whose effects are already well understood, but instead from the axion field at leading quadratic order. This will be demonstrated in Sections \ref{sec:pairproductionraterigorous}, \ref{sec:simplecases} and \ref{sec:generalsolution}. We may therefore consistently neglect the transverse excitations and restrict to the straight-string configuration $X^{\mu}(t,x) =(t,x,0,0)$. The induced metric then reduces to $h_{\alpha \beta} = \text{diag}(1,-1)$, and the action relevant for the decay-rate calculation becomes
\begin{align}
    S = \int_{\mathcal{W}} dt dx\, \left( -\kappa + \frac{1}{2} (\partial_{t} a)^{2} - \frac{1}{2} (\partial_{x} a)^{2} - \frac{1}{2} m_{a}^{2} a^{2} \right) - m_{q} \int_{\partial \mathcal{W}}ds\, .
\label{Eq:S_L}
\end{align}
As the string breaking calculation will make use of Euclidean time $\tau = it$, we also give the Euclidean action
\begin{align}
    S_{E} = \int_{\mathcal{W}} d\tau dx\, \left( \kappa + \frac{1}{2} (\partial_{\tau} a)^{2} + \frac{1}{2} (\partial_{x} a)^{2} + \frac{1}{2} m_{a}^{2} a^{2} \right) + m_{q} \int_{\partial \mathcal{W}}ds,
\label{Eq:S_E}
\end{align}
where now the line element is in Euclidean signature. It is convenient to perform an integration by parts and express this as
\begin{align}
    S_{E} = \int_{\mathcal{W}} d\tau dx\, \left(\kappa - \frac{1}{2} a \nabla^{2} a + \frac{1}{2} m_{a}^{2} a^{2}\right) + \int_{\partial \mathcal{W}} ds \left(m_{q} + \frac{1}{2} a\, \partial_{n} a \right) ,
\label{Eq:S_E_ibp}
\end{align}
where $\nabla^{2}$ is the Laplacian and $\partial_{n}$ denotes the outward normal derivative to the boundary. From this we get the Euclidean equation of motion
\begin{align}
    (\nabla^{2} - m^{2}_{a})\,a = 0 ,
\end{align}
and natural boundary condition
\begin{align}
    \partial_{n} a = 0 \ \ \text{on}\ \ \partial \mathcal{W}.
\end{align}
Finally, the equation of motion for the boundary can be obtained by varying the boundary curve $\partial \mathcal{W}$ by a normal displacement $\delta n(s)$:
\begin{align}
    \delta S_{E} = \int_{\partial \mathcal{W}} ds\, \delta n(s)\left( \kappa + \frac{1}{2} (\partial_{\tau} a)^{2} + \frac{1}{2} (\partial_{x} a)^{2} + \frac{1}{2} m_{a}^{2} a^{2} \right) - m_{q} \int_{\partial \mathcal{W}}ds\, \delta n(s) k(s),
\end{align}
where $k(s)$ is the signed curvature of the boundary worldline. Requiring the variation to vanish gives
\begin{align}
    k(s) = \frac{\kappa + \frac{1}{2} (\nabla a)^{2} + \frac{1}{2} m_{a}^{2} a^{2}}{m_{q}}\, ,
\end{align}
which simplifies upon the axion boundary condition $\partial_{n} a = 0$ to
\begin{align}
    k(s) = \frac{\kappa + \frac{1}{2} (\partial_{s} a)^{2} + \frac{1}{2} m_{a}^{2} a^{2}}{m_{q}}\, ,
\label{Eq:boundary_EOM}
\end{align}
where $\partial_s$ is the derivative with respect to arclength along the boundary.

\section{String breaking in the ground state}
\label{Sec:GroundState}

Let us first review the simplest case: the decay of an unexcited string through the nucleation of a quark-antiquark pair. Following the classic treatment of Coleman and Callan, the decay rate can be computed semi-classically using the Euclidean path integral~\cite{Coleman:1977py,Callan:1977pt}.

\begin{figure}
        \centering
        \includegraphics[width=0.9\textwidth]{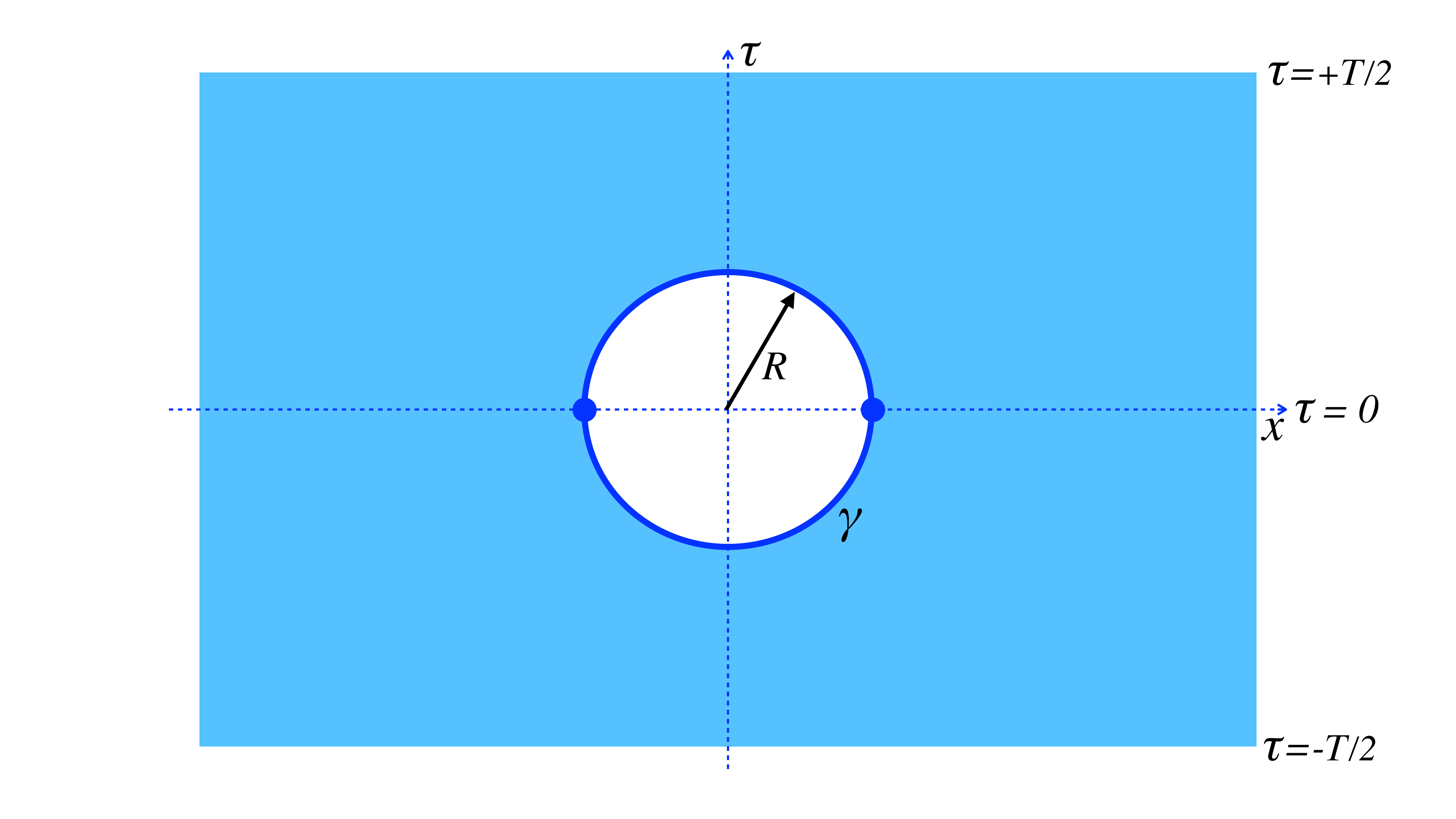}
        \caption{Bounce configuration in Euclidean spacetime. The string worldsheet $\mathcal{W}$ is shown in blue.  The hole in the worldsheet signals the breakup of the string due to the nucleation of a quark-antiquark pair which traces out the curve $\gamma$. The quarks are labelled on the timeslice $\tau = 0$ to make this clear. For a constant-tension string the critical bounce solution is a circular hole of radius $R_0=m_q/\kappa$. Usually the Euclidean time extent $T$ is taken $\to\infty$ thus preparing the ground state.}
        \label{Fig:bounce}
    \end{figure}

\begin{figure}
        \centering
        \includegraphics[width=0.9\textwidth]{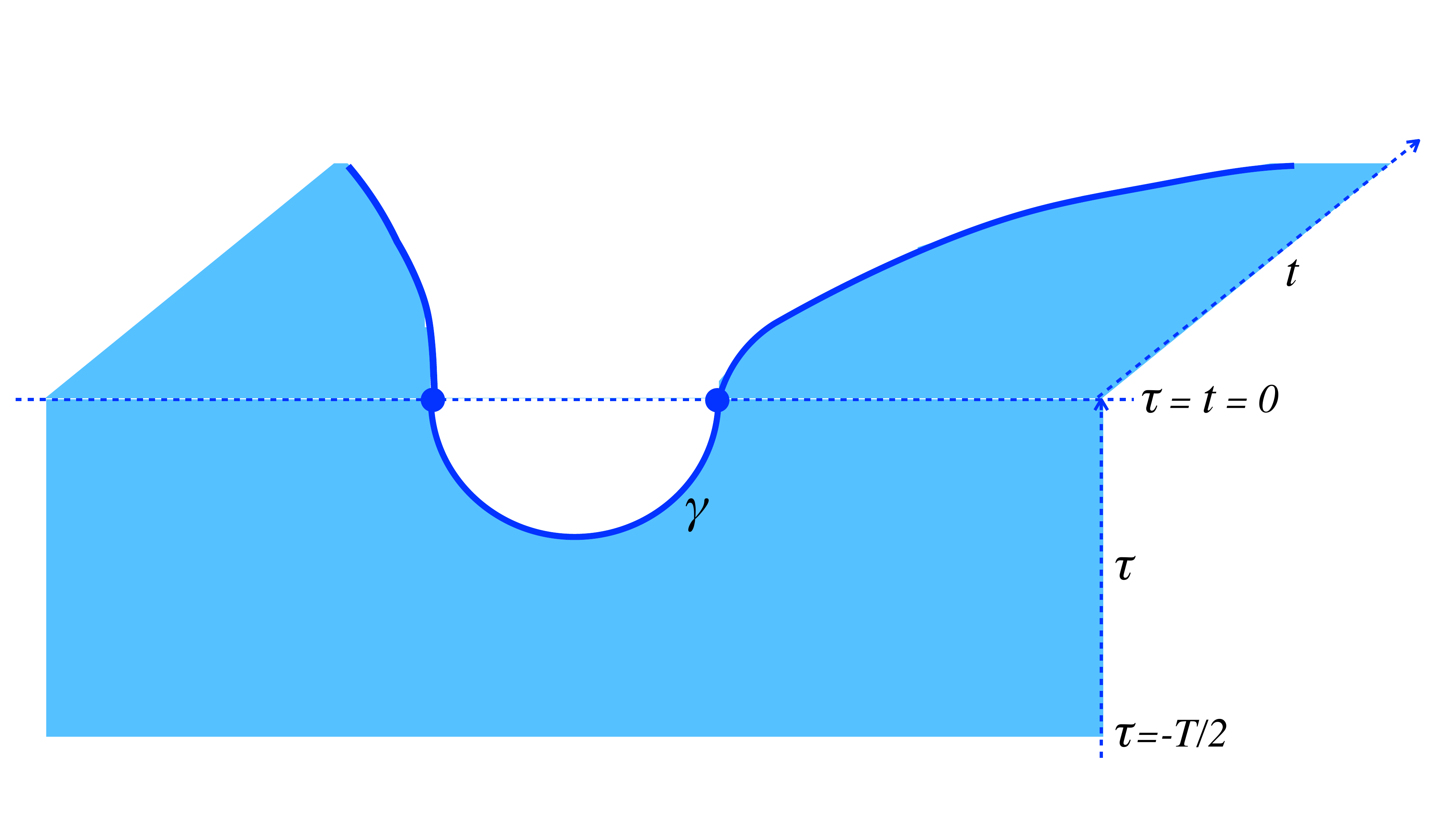}
        \caption{A mixed Euclidean-Lorentzian signature cartoon of the full decay process. At $\tau=0=t$ the state with a nucleated quark-antiquark pair, as prepared by the Euclidean-signature critical bounce solution, is analytically continued to Lorentzian signature and subsequently undergoes Lorentzian time evolution. The (anti)quark boundaries of the Lorentzian-signature string fragments are denoted by thick blue lines.}
        \label{Fig:bouncecontinuation}
    \end{figure}

In the semi-classical limit, the path integral is dominated by a saddle-point configuration known as the \emph{bounce}, shown in \Cref{Fig:bounce}. The bounce is a solution of the Euclidean equations of motion that approaches the uniform ``false vacuum" string state at Euclidean times $\tau=\pm\infty$. At the time-symmetric midpoint, $\tau = 0$, the string is broken into two segments terminated by a pointlike quark-antiquark pair separated by a distance $2R_0$. The Euclidean evolution may be viewed as preparing this configuration, which is then analytically continued to Lorentzian signature and subsequently evolves in real time, as illustrated in \Cref{Fig:bouncecontinuation}.

The resulting pair-production rate per unit string length takes the form
\begin{align}
    \frac{d \Gamma}{d \ell} = A\, e^{-B},
\label{Eq:bounce_prescription}
\end{align}
where
\begin{align}
    B\equiv S^{(b)}_{E} - S^{(0)}_{E}
\end{align}
is the bounce exponent, given by the difference between the Euclidean action evaluated on the bounce configuration $(b)$ and on the unbroken worldsheet $(0)$. The prefactor $A$ is found by performing the path integral over fluctuations about the bounce saddle. The bounce exponent must be large, $B \gg 1$, for this semi-classical treatment to be reliable.

The action difference receives contributions only from the boundary worldline $\gamma$ of the nucleated quark-antiquark pair. Using the Euclidean action Eq.~\eqref{Eq:S_E} with no axion field present, one finds
\begin{align}
    B(\gamma) = - \kappa \cdot \text{Area}(\gamma) + m_{q} \cdot \text{Length}(\gamma)~.
\end{align}
For a fixed perimeter, the enclosed area is maximised by a circle. It therefore follows that the extremal configuration is a circle of some radius $R$, for which
\begin{align}
    B(R) = 2\pi R\, m_{q}- \pi R^{2}\,\kappa.
\label{Eq:S_E_groundstate}
\end{align}
Extremising with respect to $R$ gives the critical ground-state radius
\begin{align}
    R_{0} = \frac{m_{q}}{\kappa}\,.
\end{align}
This result could also be found directly from the boundary equation of motion, Eq.~\eqref{Eq:boundary_EOM}, which requires the boundary curvature to be constant, $k(s) = \kappa/m_{q}$.

The circular solution is a true bounce: it possesses a single negative fluctuation mode corresponding to an overall rescaling of its radius, and therefore describes the decay of the metastable string state. The $O(2)$-symmetric circular hole is the direct analogue of the $O(4)$-symmetric critical bubble familiar from zero-temperature false vacuum decay in $(3+1)$ dimensions, and we will occasionally adopt the same terminology.

Evaluating the action on the critical solution gives the ground-state bounce action
\begin{align}
    B_{0} = \frac{\pi m_{q}^{2}}{\kappa},
\end{align}
and hence the decay rate
\begin{align}
    \frac{d \Gamma}{d \ell} = A \exp\left(- \frac{\pi m_{q}^{2}}{\kappa} \right).
\end{align}
This reproduces Eq.~\eqref{Eq:Lund}, thereby establishing the connection between string breaking and Schwinger pair production at leading order.\footnote{Note that throughout this work we suppress the dependence of the rate on the transverse momentum $p_\perp$ of the produced mesons. In the Lund-model literature it is often asserted that the full result is obtained by the replacement $m_q^2 \rightarrow (m_q^2 + p_\perp^2)$, motivated by a Lorentz-transformation argument. This argument is not exact, and reproduces the correct result only in the small-$p_\perp^2$ limit. A first-principles derivation of the transverse momentum distribution within string effective theory is possible, though considerably more involved; we refer the reader to a companion paper for details \cite{TransverseMomentum}.}

We can now make precise the regime of validity discussed above. The semi-classical condition $B \gg 1$ implies roughly $R_{0} \sqrt{\kappa} \gg 1$, so that the critical bubble is much larger than the string thickness and therefore remains within the domain of validity of the effective string theory. Equivalently, $R_{0} \gg 1/m_{q}$, meaning the critical bubble is much larger than the scale associated with the endpoint mass. In the language of false vacuum decay, this is precisely the thin-wall regime.

Having established the ground-state decay rate, we now turn to the effects of worldsheet axion excitations.

\section{String breaking from excited worldsheet states}
\label{sec:pairproductionraterigorous}

Axion excitations on the string worldsheet modify the above calculation of the breaking rate in several essential ways. To begin with, the worldsheet axion directly contributes to the Euclidean action Eq.~\eqref{Eq:S_E}. One might expect that the analysis of the previous section then carries through upon replacing the tension by the bulk Lagrangian density. This is \emph{not} the case. The derivation must be modified as the infinite Euclidean time evolution used previously projects onto the string ground state, exponentially suppressing any excitations. To retain the effect of excited configurations on the bounce, one must instead match the Lorentzian string configuration onto the Euclidean domain at \emph{finite} Euclidean time. This is naturally described using a complex-time contour of Schwinger–Keldysh type, with the initial state encoded through an appropriate density matrix insertion. The resulting formalism introduces additional contributions to the effective action. We outline the general framework in \Cref{Sec:Keldysh} and specialise it to the case of the worldsheet axion in \Cref{Sec:bounceexponent}.

\subsection{Schwinger-Keldysh formulation}
\label{Sec:Keldysh}

\begin{figure}
    \centering
    \includegraphics[width=0.8\textwidth]{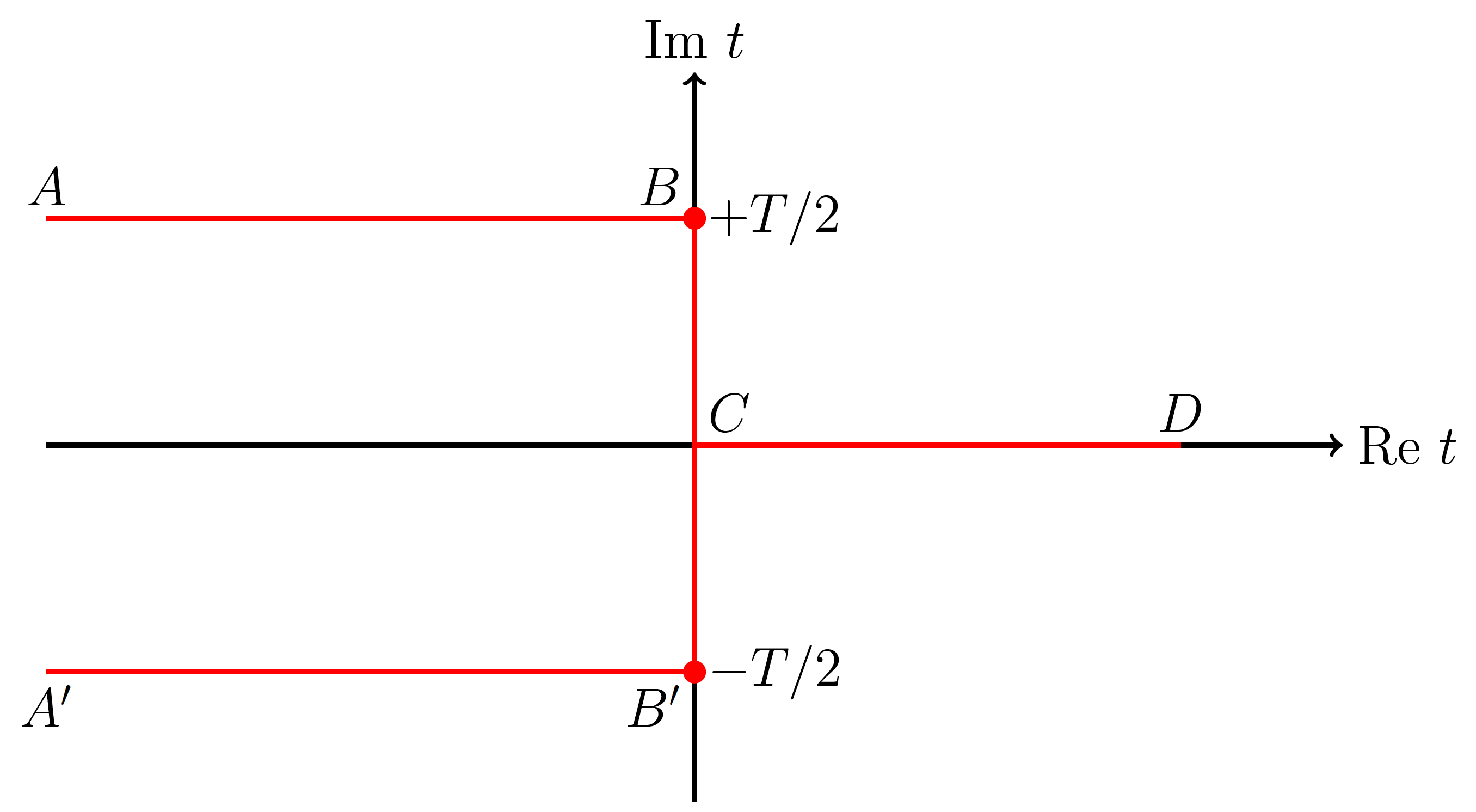}
    \caption{The contour in the complex time plane over which the double path integral is taken. Fields $\phi_{+}$ are defined on the upper branch $ABCD$, while fields $\phi_{-}$ are defined on the lower branch $A'B'CD$. Red circles indicate the density matrix insertions at $B$ and $B'$.}
    \label{Fig:SK_Contour}
\end{figure}

We follow the semi-classical path integral formalism developed by Rubakov, Son, and Tinyakov for false-vacuum decay and tunnelling from excited states ~\cite{Rubakov:1991fb,Rubakov:1992ec,Son:1995wz}. These methods have since been applied to a variety of closely related processes, including collision-induced tunnelling and inclusive transition probabilities at fixed energy or particle number~\cite{Kuznetsov:1997az,Levkov:2004tf,Levkov:2004ij,Demidov:2015bua}. In those applications, however, the excitations are carried by the same field that undergoes tunnelling. Our setup differs in two important respects. First, the tunnelling degree of freedom is the string embedding data, while the axion is a spectator excitation propagating on the worldsheet background. Second, rather than studying an inclusive ensemble at fixed energy or occupation number, we consider tunnelling in the presence of a prescribed semi-classical axion configuration. We therefore adapt the formalism accordingly.

The inclusive decay probability is expressed as a \emph{double} path integral over field configurations, including the string embedding coordinates and any additional worldsheet degrees of freedom, which we collectively denote by $\phi$. For an initial density matrix $\rho$, the properly normalized probability is
\begin{equation}
    \sigma[\rho] = \frac{1}{\cal N} \int \mathcal{D} \phi_{+} \mathcal{D}\phi_{-}\, \rho[\phi_{+},\phi_{-}] e^{iS[\phi_{+}]-iS[\phi_{-}]}.
\label{eq:sigmadef}
\end{equation}
Here $\phi_{+}$ and $\phi_{-}$ are defined on the forward and backward contours $ABCD$ and $A'B'CD$ in the complex time plane, respectively (see \Cref{Fig:SK_Contour}). The second path integral computes the complex-conjugate amplitude of the first, so \eqref{eq:sigmadef} represents a probability rather than an amplitude. The functional integral in the numerator is restricted to configurations belonging to the decayed sector at the final time $D$, namely those containing a critical bubble that has expanded and converted the false vacuum into true vacuum. The normalisation factor $\mathcal{N}$ is obtained by removing this restriction and summing over all final states.

In practice, both the numerator and denominator are evaluated in the saddle-point approximation. To exponential accuracy, the numerator is dominated by a single-bounce solution describing the nucleation of a critical hole in the worldsheet, whereas $\mathcal{N}$ is dominated by the trivial saddle in which the worldsheet remains intact and no decay occurs.

The density matrix is naturally specified at the initial points $A$ and $A'$. However, the Lorentzian segments $AB$ and $A'B'$ can be regarded as merely evolving it to $B$ and $B'$,
\begin{align}
    \rho_B = U_{AB}\rho_A U_{AB}^{\dagger}.
\end{align}
Since this evolution can be absorbed into the definition of the initial state, we may equivalently specify the density matrix directly at $B$ and $B'$. For convenience, we adopt this viewpoint and omit the segments $AB$ and $A'B'$ from the contour, taking $\rho_{B}$ as the initial density matrix.

A further simplification follows from the fact that the forward and backward contours coincide along the final Lorentzian segment $CD$. Since the fields are required to match at $D$, the path integral enforces $\phi_{+} = \phi_{-}$ throughout this segment. The contributions of $CD$ to the exponent therefore cancel identically.

The decay probability~\eqref{eq:sigmadef} can then be written schematically as
\begin{align}
    \sigma =\frac{{\rm Tr}\left[P_{\rm decay}K\rho_BK^\dagger\right]}{\mathcal{N}},
\end{align}
where $K$ denotes evolution along the Euclidean segment $BCB'$, and $P_{\rm decay}$ projects onto the decayed sector. The normalisation factor reduces similarly, $\mathcal{N} = {\rm Tr}\left[K\rho_BK^\dagger\right]$. Thus all nontrivial contributions to the decay exponent arise from the Euclidean portion of the contour. Note that the Euclidean time interval $T$ is an arbitrary parameter of the contour. Since the contour may be continuously deformed without changing the value of the path integral (provided no non-analyticities are crossed), the decay probability must be independent of $T$. We will return to this point in \Cref{sec:generalsolution}.

At leading exponential order, the decay probability is therefore determined by the difference between the Euclidean actions of two saddle configurations: the trivial saddle describing an unbroken string, and the bounce saddle describing the nucleation of a critical hole in the worldsheet.

\subsection{Excited-state correction to the bounce action}
\label{Sec:bounceexponent}

The saddle configurations are fixed by the initial density matrix insertion. To make this explicit, consider the axion field data $a(t,x)$ contained in $\phi$; we assume that the embedding degrees of freedom have already been treated separately. We are interested in a coherent initial state centred on the classical phase space data
\begin{align}
    a(t_{i},x) = \bar{a}(x), \quad \pi(t_{i},x) = \bar{\pi}(x),
\end{align}
where $\pi$ is the momentum conjugate to $a$. The coherent-state wavefunctional is obtained by shifting the false-vacuum ground-state wavefunctional\footnote{This is the QFT version of the simple harmonic oscillator ground state wavefunction $\psi_{0} \propto \exp(-m \omega q^2/2)$ with the identification $m\omega \leftrightarrow \omega_{k}$ mode-by-mode.}
\begin{align}
    \Psi_{0}[a] \propto \exp\left[ -\frac{1}{2} \int dx\, dy\, a(x) \hat{\omega}(x-y)a(y)\right].
\end{align}
Here, $\hat{\omega}$ is formally defined through the operator equation $\hat{\omega} = \sqrt{-\partial_{x}^{2} + m^{2}}$, or more precisely
\begin{align}
    \hat{\omega}(x-y) = \int \frac{dp}{2\pi} \,\omega_{p} e^{ip(x-y)}, \qquad \omega_{p} = \sqrt{p^{2} + m^{2}}\, .
\end{align}
The coherent-state wavefunctional is then
\begin{align}
    \Psi_{\bar{a},\bar{\pi}}[a] &= \exp\left[i \int dx\, \bar{\pi}\, a\right]\exp\left[i \int dx\, \bar{a}\, \pi\right] \Psi_{0}[a]\\
        &\propto \exp\left[ -\frac{1}{2} \int dx\, dy\, [a(x)-\bar{a}(x)] \hat{\omega}(x-y)[a(y)-\bar{a}(y)] + i \int dx\, \bar{\pi}(x) a(x) \right]. \nonumber
\end{align}
The density matrix in this state is
\begin{align}
    \rho = \ket{\Psi_{\bar{a},\bar{\pi}}} \bra{\Psi_{\bar{a},\bar{\pi}}},
\end{align}
so the double path integral contains
\begin{align}
    \rho[a_{+},a_{-}] = \bra{a_{+}} \rho \ket{a_{-}} = \Psi_{\bar{a},\bar{\pi}}[a_{+}]\Psi_{\bar{a},\bar{\pi}}^{*}[a_{-}],
\end{align}
with the asterisk denoting complex conjugation. Note that here $a_{+}$ is to be evaluated at the initial time $t_{i}$ and $a_{-}$ at the complex conjugate time $t_{i}^{*}$. With the density matrix insertion, then, the decay probability takes the form
\begin{align}
    \sigma = \frac{1}{\mathcal{N}} \int \mathcal{D}a_{+}\mathcal{D}a_{-}\, e^{i S_{\text{SK}}[a_{+},a_{-}]}
\end{align}
with the Schwinger-Keldysh effective action defined as
\begin{align}
\label{Eq:Seff_phi}
     iS_{\text{SK}}[a_{+},a_{-}] &= iS[a_{+}] - iS[a_{-}] \\
     &- \frac{1}{2} \int dx dy\, [a_{+}(t_{i},x) - \bar{a}(x)] \hat{\omega}(x-y) [a_{+}(t_{i},y) - \bar{a}(y)] \nonumber\\
     &- \frac{1}{2} \int dx dy\, [a_{-}(t_{i}^{*},x) - \bar{a}^{*}(x)] \hat{\omega}(x-y) [a_{-}(t_{i}^{*},y) - \bar{a}^{*}(y)] \nonumber\\
     &+ i \int dx\, \bar{\pi}(x) a_{+}(t_{i},x) - i \int dx\, \bar{\pi}^{*}(x) a_{-}(t_{i}^{*},x). \nonumber
\end{align}
Note the density matrix insertion contributes only boundary terms at the insertion points and so does not affect the bulk equations of motion. Given the axion action Eq.~\eqref{Eq:S_L}, requiring $S_{\text{SK}}$ be stationary at the insertion points gives the boundary conditions
\begin{align}
\label{Eq:phi_BC1}
    \partial_{t}a_{+}(t_{i},x) &= \bar{\pi}(x) + i \sqrt{-\partial_{x}^{2} + m^{2}}\left[a_{+}(t_{i},x) - \bar{a}(x)\right], \\\label{Eq:phi_BC2}
    \partial_{t}a_{-}(t_{i}^{*},x) &= \bar{\pi}^{*}(x) - i \sqrt{-\partial_{x}^{2} + m^{2}}\left[a_{-}(t_{i}^{*},x) - \bar{a}^{*}(x)\right],
\end{align}
where the operator is to be interpreted in terms of its Fourier representation as above.

As discussed above, only the Euclidean segment of the contour contributes to the tunnelling exponent, and we place the density matrix insertion at $B$ and $B'$ for convenience, corresponding to $t_{i} = iT/2$. The problem then reduces entirely to evolution along the Euclidean interval $-T/2 \leq \tau \leq T/2$. Since the fields on the two branches must coincide at the matching point $t=0$, the saddle may be described using a single Euclidean field configuration $a(\tau,x)$ defined on the full interval. The boundary values are identified as $a_{+}(t_{i},x) = a(\tau = -T/2,x)$ and $a_{-}(t_{i}^{*},x) = a(\tau = +T/2,x)$. After Wick rotation, the effective action Eq.~\eqref{Eq:Seff_phi} becomes
\begin{align}
     -S_{\text{SK},E}[a] = &-S_{E}[a]\\
     &- \frac{1}{2} \int dx dy\, [a(-T/2,x) - \bar{a}(x)] \hat{\omega}(x-y) [a(-T/2,y) - \bar{a}(y)] \nonumber\\
     &- \frac{1}{2} \int dx dy\, [a(+T/2,x) - \bar{a}^{*}(x)] \hat{\omega}(x-y) [a(+T/2,y) - \bar{a}^{*}(y)] \nonumber\\
     &+ i \int dx\, \bar{\pi}(x) a(-T/2,x) - i \int dx\, \bar{\pi}^{*}(x) a(+T/2,x), \nonumber
\end{align}
and the boundary conditions Eqs.~\eqref{Eq:phi_BC1} and \eqref{Eq:phi_BC2} are expressed as
\begin{align}
    \label{eq:insertioneqns1}
    \partial_{\tau}a(-T/2,x) &= -i\bar{\pi}(x) +  \sqrt{-\partial_{x}^{2} + m^{2}_{a}}\left[a(-T/2,x) - \bar{a}(x)\right], \\
    \label{eq:insertioneqns2}
    \partial_{\tau}a(+T/2,x) &= -i\bar{\pi}^{*}(x) -  \sqrt{-\partial_{x}^{2} + m^{2}_{a}}\left[a(+T/2,x) - \bar{a}^{*}(x)\right].
\end{align}
The bulk contribution of the axion to the Euclidean action Eq.~\eqref{Eq:S_E_ibp} vanishes on-shell. However, because the coherent-state insertion modifies the boundary conditions at $\tau = \pm T/2$ the on-shell action receives a non-trivial boundary contribution:
\begin{align}
    S_{a,E}[a] = \frac{1}{2} \int_{\partial \mathcal{W}} ds \, a\, \partial_{n} a = \frac{1}{2} \int dx \, a\, \partial_{\tau} a_{|\tau = +T/2} - \frac{1}{2} \int dx \, a\, \partial_{\tau} a_{|\tau = -T/2}.
\end{align}
This is true even in the bounce geometry, as the axion is subject to the natural boundary condition $\partial_{n} a = 0$ on the hole boundary.

We will determine the modification to the bounce exponent using the following procedure. First, we specify a classical Lorentzian background configuration on the unbroken string, $a = a^{(0)}(t,x)$, in which we want to find the nucleation rate at $(t,x) = (0,0)$. This background translates to coherent-state initial data\footnote{The
coherent-state labels $\bar a,\bar\pi$ appearing in the density-matrix insertion are the analytic continuations to the contour insertion point $t_i= iT/2$. Moreover, since the string itself is absent at $(t,x) = (0,0)$ upon nucleation, references to field values there should be understood as referring instead to the Lorentzian no-bounce data at the candidate nucleation point.}
\begin{align}
    \bar{a}(x) \equiv a^{(0)}(t_{i},x), \quad \bar{\pi}(x) \equiv \partial_{t} a^{(0)}(t_{i},x).
\end{align}
For the trivial worldsheet geometry, the Euclidean saddle is simply the analytic continuation of the given Lorentzian profile,
\begin{align}
    a_{E}^{(0)}(\tau,x) = a^{(0)}(-i\tau,x).
\end{align}
Indeed, given the physical condition that the Lorentzian solution is real on the real time axis, this automatically satisfies
\begin{align}
    a_{E}^{(0)}(-T/2,x) = \bar{a}(x) \quad \text{and} \quad  a_{E}^{(0)}(+T/2,x) = \bar{a}^{*}(x),
\end{align}
as well as
\begin{align}
    \partial_{\tau}a_{E}^{(0)}(-T/2,x) = -i\bar{\pi}(x) \quad \text{and} \quad  \partial_{\tau}a_{E}^{(0)}(+T/2,x) = -i\bar{\pi}^{*}(x),
\end{align}
and therefore the boundary conditions are satisfied.

On the bounce geometry, the solution differs from the trivial continuation by a correction $\Delta_{a}$ induced by the hole in the worldsheet:
\begin{align}
    a_{E}^{(b)}(\tau,x) = a_{E}^{(0)}(\tau,x) + \Delta_{a}(\tau,x).
\end{align}
The correction must satisfy the bulk equations of motion
\begin{align}
    (\nabla^{2} - m_{a}^{2}) \Delta_{a} = 0,
\label{Eq:Delta_EOM}
\end{align}
together with boundary conditions
\begin{align}
\label{Eq:Delta_BC1}
    \partial_{\tau}\Delta_{a}(-T/2,x) &= + \sqrt{-\partial_{x}^{2} + m^{2}_{a}} \,\Delta_{a} (-T/2,x), \\
\label{Eq:Delta_BC2}
    \partial_{\tau}\Delta_{a}(+T/2,x) &= - \sqrt{-\partial_{x}^{2} + m^{2}_{a}} \,\Delta_{a} (+T/2,x),
\end{align}
which follow from the linearity of Eqs.~\eqref{eq:insertioneqns1}-\eqref{eq:insertioneqns2}. In addition, imposing natural boundary conditions requires
\begin{align}
    \partial_{n} \Delta_{a} = -\partial_{n} a^{(0)} \quad \text{on the hole boundary.}
\label{Eq:Delta_discBC}
\end{align}
Equations \eqref{Eq:Delta_EOM}-\eqref{Eq:Delta_discBC} define a well-posed boundary value problem for $\Delta_{a}$. We will solve for $\Delta_{a}$ in two cases of interest shortly. 

With these fields, the axion contribution to the bounce exponent $B = B_{0} + \delta B_{a}$ is
\begin{align}
\delta B_{a} = S_{\text{SK},E}\left[a^{(0)}_{E} + \Delta_{a}\right] - S_{\text{SK},E}\left[a^{(0)}_{E}\right].
\end{align}
Substituting the on-shell action and boundary conditions, all terms quadratic in $\Delta_{a}$ cancel, while the remaining linear terms combine to give 
\begin{align}
\label{Eq:DeltaS_effE}
    \delta B_{a} = \frac{1}{2} \int dx \left(a^{(0)}_{E} \partial_{\tau} \Delta_{a} -  \Delta_{a} \partial_{\tau} a^{(0)}_{E}\right)_{|\tau = +\frac{T}{2}} - \frac{1}{2} \int dx \left(a^{(0)}_{E} \partial_{\tau} \Delta_{a} -  \Delta_{a} \partial_{\tau} a^{(0)}_{E}\right)_{|\tau = -\frac{T}{2}}\, .
\end{align}
This expression is exact. However, obtaining an exact expression for $\Delta_{a}$ is highly non-trivial as the bounce geometry and the axion profile must be determined self-consistently: the axion solution depends on the geometry of the hole boundary, but this itself is subject to axion backreaction. In practice, we determine both $\Delta_{a}$ and the hole geometry perturbatively in the amplitude of the initial axion excitation. Before moving on to the computation of $\Delta_{a}$ we therefore first turn to the determination of the correct bounce boundary.

\subsection{Backreaction on the bounce geometry}\label{sec:shapeofbounce}

With the axion field present, the solution to the endpoint equation of motion is only approximately a circle. The curvature of the boundary is determined by the equation of motion Eq.~\eqref{Eq:boundary_EOM}
\begin{align}
    k(s) = \frac{\kappa + \varepsilon(s)}{m_{q}}\, ,
\end{align}
where
\begin{align}
    \varepsilon(s) = \frac{1}{2} (\partial_{s} a)^{2} + \frac{1}{2} m_{a}^{2} a^{2}.
\label{Eq:epsilon_def}
\end{align}
Here the axion field is that of the bounce configuration, $a = a^{(0)}_{E} + \Delta_{a}$.

If $\varepsilon$ were constant along the boundary, the solution would simply be a circle as before. We therefore treat the true bounce as a small deformation away from a circle and parameterise the boundary by $r(\theta) = R_{\star} + \xi(\theta)$.
To leading order in the deformation, we may evaluate Eq.~\eqref{Eq:epsilon_def} as
\begin{align}
    \varepsilon = \frac{1}{2R^{2}_{\star}} (\partial_{\theta} a)^{2} + \frac{1}{2} m_{a}^{2} a^{2}.
\end{align}
The curvature of the deformed boundary is
\begin{align}
    k(\theta) = \frac{r^{2} + 2 \xi'^{2} - r \xi''}{(r^{2} + \xi'^{2})^{3/2}}\simeq \frac{1}{R_{\star}} - \frac{\xi + \xi''}{R^{2}_{\star}}\, ,
\end{align}
to first order in $\xi$.
It is convenient to expand the deformation in a Fourier basis,
\begin{align}
    \xi(\theta) = \sum_{n=2}^{\infty} \left[\xi_{n}^{(c)} \cos(n \theta) + \xi_{n}^{(s)} \sin(n \theta) \right],
\end{align}
and similarly
\begin{align}
    \varepsilon(\theta) = \varepsilon_{0} + \sum_{n=2}^{\infty} \left[\varepsilon_{n}^{(c)} \cos(n \theta) + \varepsilon_{n}^{(s)} \sin(n \theta) \right].
\end{align}
Substituting the Fourier expansions into the equation of motion yields the solution
\begin{align}
    \xi_{n}^{(c)} = \frac{R^{2}_{\star}}{m_{q}} \frac{\varepsilon_{n}^{(c)}}{n^{2}-1}, \quad \xi_{n}^{(s)} = \frac{R^{2}_{\star}}{m_{q}} \frac{\varepsilon_{n}^{(s)}}{n^{2}-1},
\label{Eq:xi_solution}
\end{align}
where the modified average bounce radius depends on the constant mode $\varepsilon_{0}$,
\begin{align}
R_{\star} = \frac{m_{q}}{\kappa + \varepsilon_{0}}.
\label{Eq:R_modified}
\end{align}
Note that $n=1$ modes correspond to translations of the circle and do not represent physical deformations of the bounce profile, and so are not included.\footnote{For a homogeneous string, they are collective-coordinate shifts. For a genuinely inhomogeneous axion background, they move the preferred nucleation centre; they do not affect the leading local exponent when the background varies slowly over $R_{\star}$. For approximately homogeneous and time-independent backgrounds where the variation of the axion field is small over the scale $R_{\star}$ we have approximate zero-mode collective coordinates corresponding to local worldsheet time and space translations, and thus in this case we may think in terms of a local string decay rate per unit length if we so wish.}

The deformation of the bounce boundary has several effects. First, the length and enclosed area of the curve $\gamma$ are no longer exactly $2\pi R_{\star}$ and $\pi R^{2}_{\star}$. However, these corrections do not contribute at leading order to the bounce action because the circular solution with constant $\varepsilon$ is an extremum. As we will be working to lowest order in the excitations of the axion field, the \emph{geometric contribution} to our bounce action will therefore still have the circular form Eq.~\eqref{Eq:S_E_groundstate}, although with a radius modified by the axion field.

Secondly, the Neumann boundary condition that determines $\Delta_{a}$ depends on the shape of the boundary. Specifically, the normal derivative differs from the radial derivative by terms proportional to the deformation, 
\begin{align}
    \partial_{n} = \frac{1}{\sqrt{1+\xi'^{2}/r^{2}}} \left( \partial_{r} - \frac{\xi'}{r^{2}} \partial_{\theta} \right) = \partial_{r} - \frac{\xi'}{R^{2}_{\star}} \partial_{\theta} + \mathcal{O}\!\left(\frac{\xi'^{2}}{R^{2}_{\star}}\right).
\end{align}
Since $\Delta_{a}$ is already computed perturbatively in the axion data, the correction from replacing $\partial_{n}$ by $\partial_{r}$ contributes only at higher order. Thus, to obtain the leading term in $\Delta_{a}$, it is consistent to evaluate the boundary condition on the unperturbed circular bounce.

Finally, since the bounce is not exactly circular, there will generically be non-zero time derivatives when matching onto Lorentzian evolution. As we will soon see, this leads to initial longitudinal velocities of the quarks at nucleation. The physics of the bounce boundary shape deformation and its consequences for energy and longitudinal momentum conservation are discussed in Sections \ref{sec:boundarydeformation} and \ref{sec:energyconservation}.

\section{Massless axion background}\label{sec:simplecases}

It is instructive to first consider a simplified case which is both sufficient to derive the form of the leading axion corrections to the decay exponent, and illuminates other features of the solution. Consider the case where the worldsheet axion is massless $m_a=0$ and take the initial Lorentzian axion field profile to be simply
\begin{align}
    a^{(0)}(t,x) = a_{x} x + a_{t} t,
\end{align}
with $a_x$ and $a_{t}$ constant real parameters.\footnote{Such a globally linear profile is not a normalisable state on the infinite line; it should be understood as the low-$k$ limit of a wavepacket.} Then the initial coherent-state data at the insertion points is $\bar{a}(x)=a_x x + ia_t T/2$, and $\bar{\pi}(x)=a_t$. The corresponding Euclidean background profile is the analytic continuation
\beq
a_E^{(0)}(\tau,x)=a_x x-ia_t \tau,
\eeq
which satisfies $\nabla^2 a_E^{(0)} =0$. Following the procedure described in \Cref{Sec:bounceexponent}, we write the bounce solution as
\beq
a^{(b)}_{E}=a^{(0)}_{E} +\Delta_{a},
\eeq
with the correction also satisfying $\nabla^2\Delta_{a}=0$. In this massless case the boundary conditions at $\tau=\pm T/2$, Eqs.~\eqref{Eq:Delta_BC1}-\eqref{Eq:Delta_BC2}, reduce to
\beq
\partial_\tau\Delta_{a}(-T/2,x)=+|\partial_x|\,\Delta_{a}(-T/2,x)~,\qquad
\partial_\tau\Delta_{a}(+T/2,x)=-|\partial_x|\,\Delta_{a}(+T/2,x)~,
\eeq
while on the trial circular hole at $r=R$ (with the value of $R$ to be determined), we require
\beq
\partial_r\Delta_{a}\big|_{r=R} =-\partial_r a_E^{(0)}\big|_{r=R}~.
\eeq
Together these are just the strip boundary conditions coming from the density-matrix insertion with the leading Neumann condition on the bounce boundary. 

By inspection the solution for the correction $\Delta_{a}$ satisfying all these conditions is simply
\beq
\Delta_{a}(\tau,x)=a_x\,\frac{R^2 x}{x^2+\tau^2} - i a_t\,\frac{R^2 \tau}{x^2+\tau^2}~,
\eeq
or, equivalently in polar coordinates on the Euclidean worldsheet, $x=r\cos\theta,\ \tau=r\sin\theta$,
\beq
a_E^{(0)}=r\bigl(a_x\cos\theta-i a_t\sin\theta\bigr),\qquad
\Delta_{a} =
\left(\frac{R^2}{r}\right)\bigl(a_x\cos\theta-ia_t\sin\theta\bigr).
\label{eq:simpleDelta}
\eeq
This is both harmonic and enforces the Neumann condition at $r=R$ correctly.  One also finds that the density-matrix insertion boundary conditions at $\tau=\pm T/2$ are satisfied upon using the Fourier representations
\bea
\frac{x}{x^2+\tau^2} & = & \int\frac{dp}{2\pi}\, \bigl(-i\pi\,\mathrm{sgn}(p)\bigr)e^{ipx-|\tau p|}~,\\
\frac{\tau}{x^2+\tau^2} & = & \int\frac{dp}{2\pi}\, \bigl(\pi\,\mathrm{sgn}(\tau)\bigr)e^{ipx-|\tau p|}~.
\eea
Thus $a^{(b)}_{E}(\tau,x)=a_E^{(0)}+\Delta_{a}$ given by Eq.~\eqref{eq:simpleDelta} is the exact solution of the strip problem for this massless axion test case, \emph{assuming a circular bubble boundary}.  In fact, as we will soon see, because of the angular dependence of Eq.~\eqref{eq:simpleDelta} the bubble boundary is distorted away from circularity, but this distortion causes a change to $\Delta_{a}$ which is higher order in the parameters $a_x$ and $a_t$.  The above solution for $\Delta_{a}$ will be sufficient to find the modified bounce action including $O(a_{x,t}^2)$ corrections, as well as other properties of the string decay.

\begin{figure}
    \centering
    \includegraphics[width=1\textwidth]{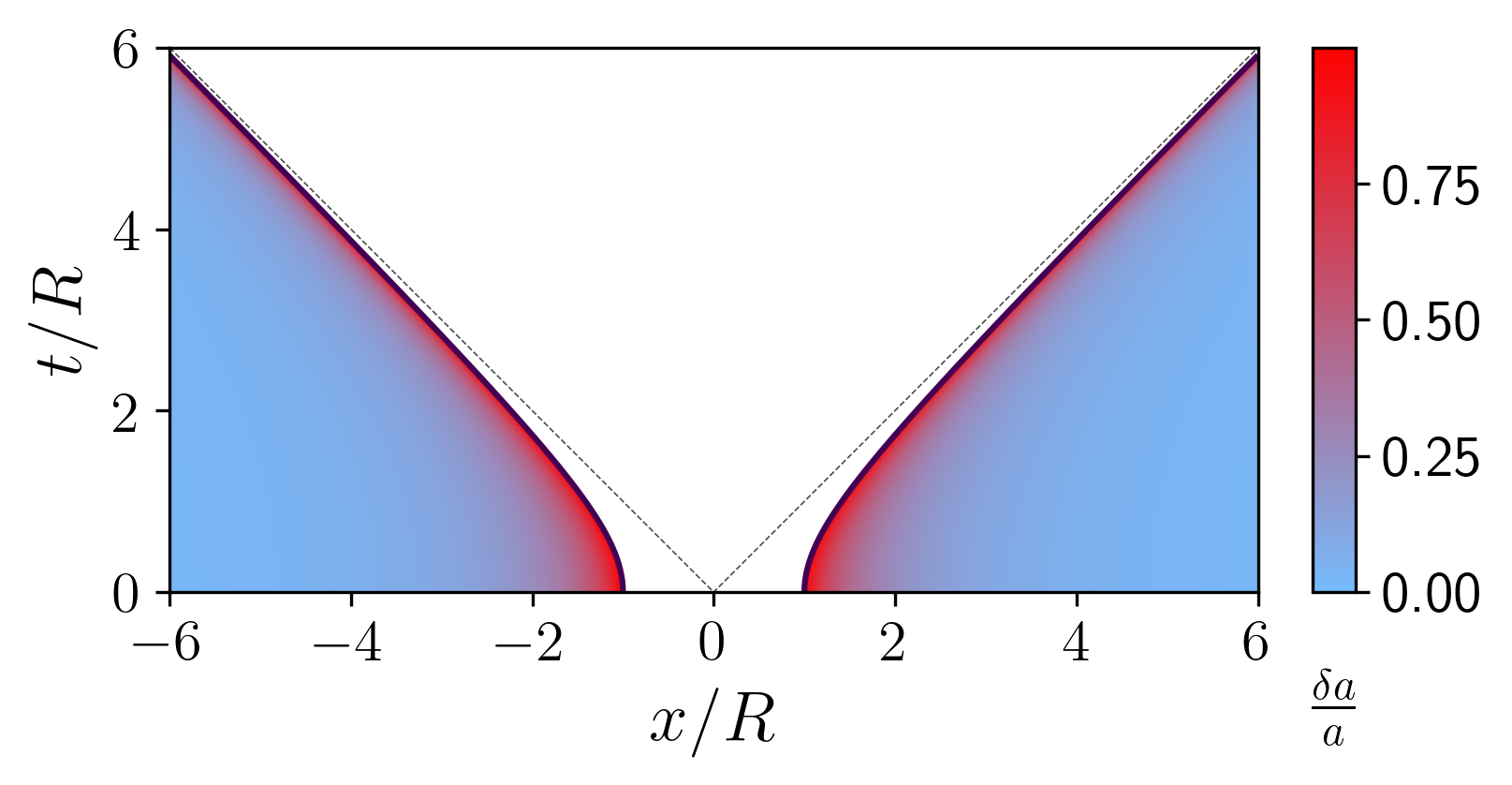}
    \caption{Fractional change in the massless axion field profile on the Lorentzian-signature string worldsheet after bubble nucleation, $(a^{(b)} - a^{(0)})/a^{(0)}$. The hyperbolic boundaries are the worldlines of the quark-antiquark pair undergoing constant proper acceleration due to the string tension. These trajectories are obtained by analytic continuation of the circular bounce solution, neglecting axion backreaction. At late times they asymptotically approach the lightcone, indicated by the dashed diagonal lines.}
    \label{Fig:axion_change}
\end{figure}

From Eq.~\eqref{eq:simpleDelta} the Lorentzian-signature axion field outside the nucleated hole is
\beq
a^{(b)}(t,x)=(a_x x+ a_t t) \left(1+\frac{R^2}{x^2-t^2}\right)~
\label{eq:axion_Lorentzian}
\eeq
for the surviving string segments $|x|\geq \sqrt{R^2 +t^2}$.
So we find that the influence of the bubble nucleation event \emph{extends beyond the removed portion of the
string}, in this case of a massless axion, only falling off quadratically over distance scales of order $R$. The fractional change in the axion field on the string worldsheet after nucleation is shown in Figure~\ref{Fig:axion_change}. 
The fractional distortion is largest near the accelerating endpoint trajectories, where it approaches order unity.  Note that the formal lightcone singularity lies outside the physical post-nucleation worldsheet.
For the realistic situation of a massive axion, as we discuss in \Cref{sec:generalsolution}, the influence of bubble nucleation extends a distance $\sim 1/m_a$ beyond the bubble boundary curve with an exponential drop-off beyond.

The contribution of the axion initial data to the bounce exponent, Eq.~\eqref{Eq:DeltaS_effE}, can now be evaluated using the explicit forms of $\Delta_{a}$ and $a_E^{(0)}$ given in Eq.~\eqref{eq:simpleDelta}. The leading order contribution, which is reliably captured by the circular-bubble approximation, is simply
\beq
\delta B_{a} = -\pi R^2\bigl(a_x^2-a_t^2\bigr).
\label{Eq:B_a_massless}
\eeq
It is a non-trivial check on our method that this result is independent of the contour parameter $T$. This independence is required as the contour may be deformed in the complex-time plane without encountering any singularities, provided $T/2\geq R$ (we briefly return to the origin of this constraint in \Cref{sec:generalsolution}).

Including the geometric contributions from the quark worldline, the total bounce exponent as a function of $R$ is then
\beq
B(R) = 2\pi R\, m_q -\pi R^2\left(\kappa+a_x^2-a_t^2\right).
\eeq
The structure of this correction suggests that the axion background induces a local shift of the string tension, $\kappa \to \kappa +a_x^2-a_t^2$. As we show more generally in \Cref{sec:generalsolution}, this interpretation extends to the massive theory, where the effective tension becomes
\beq
\kappa_{\rm eff} = \kappa+a_x^2-a_t^2 + \frac12 m_a^2 a^2 ,
\label{Eq:Keff}
\eeq
or equivalently
\beq
\kappa_{\rm eff}(t,x) =  \kappa  + (\partial_x a^{(0)})^2 - (\partial_t a^{(0)})^2 + \frac12 m_a^2 (a^{(0)})^2.
\label{eq:effective_tension}
\eeq
This correction is a worldsheet Lorentz scalar and is therefore frame-independent. The mass-term contribution in Eq.~\eqref{eq:effective_tension} can be understood from the form of the string action, as a constant uniform shift of the value of the axion field is indistinguishable from a change in $\kappa$ (at least in the limit $m_a R \ll 1$). More striking is the factor-of-two enhancement of the derivative contribution relative to the mass term, while the relative sign between spatial and temporal gradients is fixed by Lorentz invariance. The origin of the differing coefficients will become clear in \Cref{sec:generalsolution}.

Assuming the axion field varies slowly on the scale of the bounce, the exponent can be written more generally as
\beq
B(R) = 2\pi R\, m_q -\pi R^2\kappa_{\rm eff}.
\eeq
Extremising with respect to $R$ yields the critical bubble radius
\beq
R_\star =\frac{m_q}{\kappa_{\rm eff}}\, ,
\label{Eq:R_kappaeff}
\eeq
and hence the bounce action
\begin{align}
    B = \frac{\pi m_{q}^{2}}{\kappa_{\rm eff}}\, .
\end{align}
The local string-breaking rate per unit length is therefore
\begin{align}
    \frac{d \Gamma}{d \ell} \propto \exp\left(- \frac{\pi m_{q}^{2}}{\kappa_{\text{eff}}(t,x)} \right).
\label{Eq:decay_rate_eff}
\end{align}
This is one of the primary results of our analysis. Worldsheet axion excitations locally modify the effective string tension and thereby exponentially enhance or suppress string breaking. In particular, the mass term and spatial-gradient term exponentially enhance the decay rate, while the temporal kinetic term suppresses it. This result holds provided the axion-induced corrections to the exponent are perturbatively small.

\subsection{Bounce geometry and endpoint kinematics}\label{sec:boundarydeformation}

Returning to the simple massless axion case, non-vanishing gradients $a_x$ and $a_t$ generally deform the circular trial bounce boundary. To determine the leading correction to the bubble profile, we evaluate the arc-length derivative of the total axion field Eq.~\eqref{eq:simpleDelta} on the undeformed circular contour $r=R$:
\beq
\partial_s a_E^{(b)}=\frac1R\partial_\theta a_E^{(b)}
=2\bigl(-a_x\sin\theta-ia_t\cos\theta\bigr). 
\eeq
The contribution to the curvature Eq.~\eqref{Eq:epsilon_def} is therefore
\beq
\varepsilon(\theta)=\frac12\left(\partial_s a_{E}^{(b)}\right)^2 =
a_x^2-a_t^2-(a_t^2+a_x^2)\cos 2\theta+2i a_x a_t \sin 2\theta.
\eeq
In the notation of \Cref{sec:shapeofbounce}, this corresponds to
\beq
\varepsilon_0=a_x^2-a_t^2, \quad \varepsilon_2^{(c)}=-(a_t^2+a_x^2), \quad \varepsilon_2^{(s)}= 2i a_x a_t\, .
\eeq
Using Eqs.~\eqref{Eq:xi_solution} and \eqref{Eq:R_modified}, the resulting leading-order deformation of the bounce boundary is
\begin{align}
    r(\theta) = R_{\star} - \frac{R^2_{\star}}{3m_q}(a_t^2 + a_x^2) \cos 2\theta + i\frac{2R^2_{\star}}{3m_q}a_x a_t\sin 2\theta,
\label{Eq:bubble_deformed}
\end{align}
with average radius
\beq
R_{\star} = \frac{m_q}{\kappa+a_x^2-a_t^2}\, ,
\eeq
which agrees with the $m_a=0$ limit of Eq.~\eqref{Eq:R_kappaeff}.

\begin{figure}
    \centering
    \includegraphics[width=1.05\textwidth]{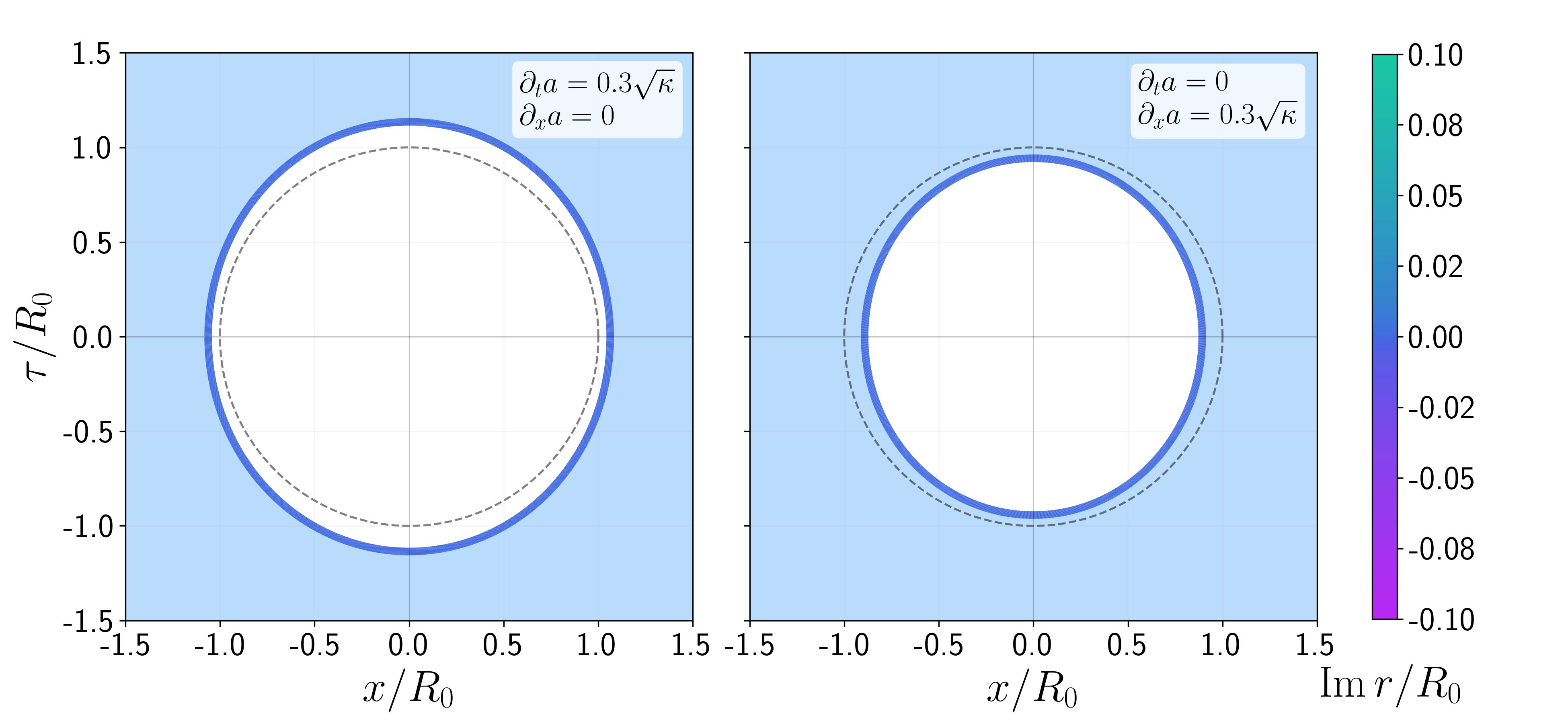}
    \caption{Leading-order deformation of the bubble profile induced by axion field gradients for fields varying purely in time (left) and space (right). In these special cases the bubble is purely real. The string worldsheet is shaded light blue (unrelated to the colour scale). The dashed curve indicates the ground-state circular bounce solution with radius $R_{0} = m_{q}/\kappa$.}
    \label{Fig:bounce_deformation_real}
\end{figure}

\begin{figure}
    \centering
    \includegraphics[width=1.05\textwidth]{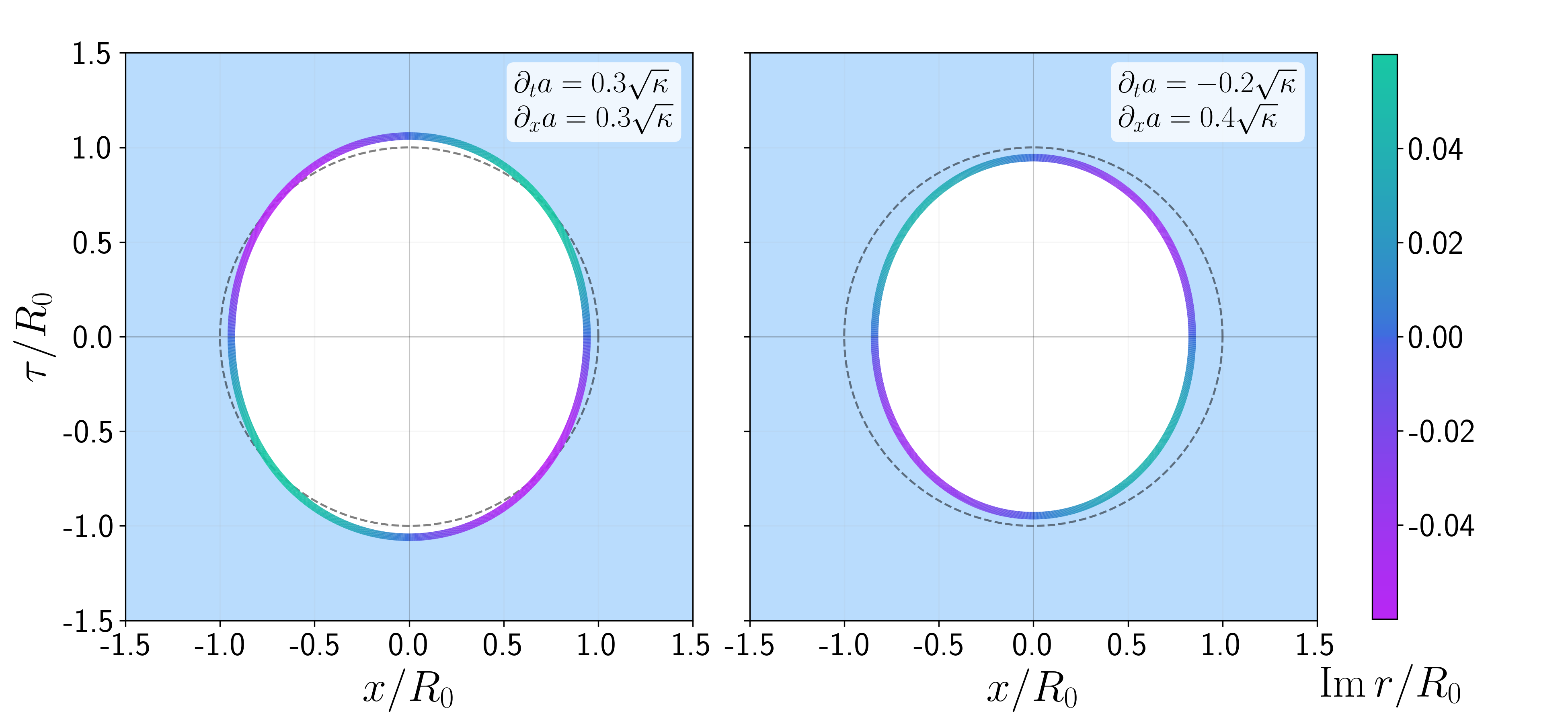}
    \caption{Leading-order deformation of the bubble profile for massless axion fields varying in both space and time. Since the solution \eqref{Eq:bubble_deformed} is complex in this case, the plots show the projection of $r(\theta)$ onto the real $(x,\tau)$ plane, while the colour scale indicates its imaginary component. The string worldsheet is shaded light blue (unrelated to the colour scale). The dashed curve indicates the ground-state circular bounce solution with radius $R_{0} = m_{q}/\kappa$.}
    \label{Fig:bounce_deformation_complex}
\end{figure}

Several features of this solution are worth noting. If either $a_x = 0$ or $a_t = 0$, the deformation remains purely real, rescaling the size of the bubble and elongating it along the $\tau$ direction, as shown in \Cref{Fig:bounce_deformation_real}. When both gradients are present, however, the coefficient of the $\sin 2\theta$ mode is purely imaginary, so the \emph{Euclidean bounce itself is complex-valued}. Such complex saddles are expected for vacuum decay in excited backgrounds and have appeared previously in studies of tunnelling from non-trivial initial states, in both quantum mechanical~\cite{Draper:2023fkz,Janssen:2026ybl} and field-theoretic examples~\cite{Rubakov:1992ec,Son:1995wz,Kuznetsov:1997az,Levkov:2004tf}. The resulting complex deformation is shown in \Cref{Fig:bounce_deformation_complex}.

Despite the Euclidean solution being complex, the physically relevant Lorentzian continuation is \emph{real}. In particular, the nucleation points at $\theta=0, \pi$ are purely real with purely real time derivatives, and therefore define sensible initial conditions for the subsequent Lorentzian-time evolution. The imaginary Euclidean component manifests itself after analytic continuation as a non-zero initial longitudinal velocity for the quark endpoints.

\begin{figure}
    \centering
    \includegraphics[width=1\textwidth]{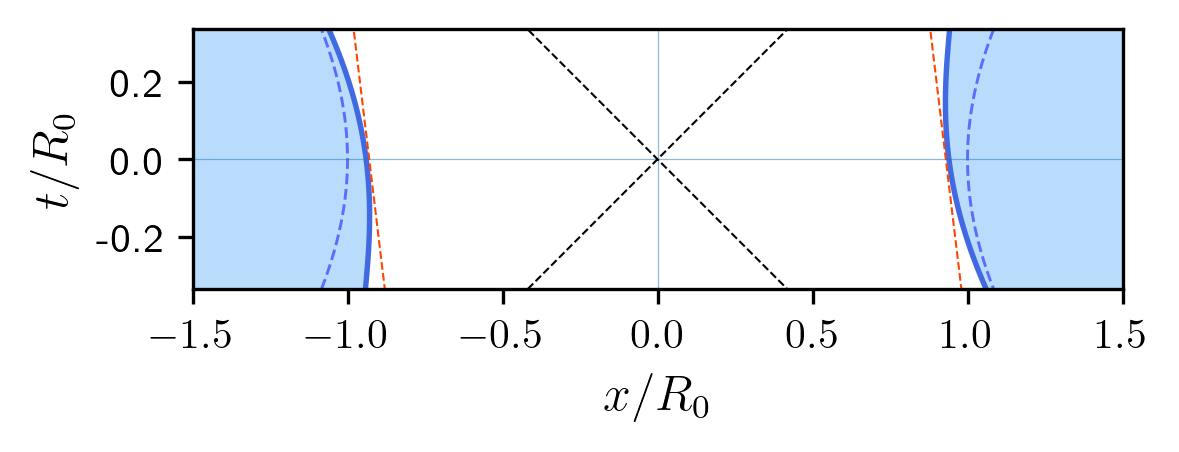}
    \caption{Lorentzian continuation of the deformed bounce for $\partial_{t} a = \partial_{x} a = 0.3 \sqrt{\kappa}$. The nucleated quark trajectories are shown as the solid blue curves bounding the string worldsheet (shaded light blue). For comparison, the ground-state solution $x^{2} - t^{2} = R_{0}^{2}$ is shown as the dashed blue curve. The modified trajectories emerge at $t=0$ with equal nonzero longitudinal velocities, signified by the dashed orange tangent lines. The dashed diagonal lines mark the lightcone. Unphysical large distortions of the quark trajectories near the lightcone have been omitted from the plot.}
    \label{Fig:bounce_deformation_Lorentzian}
\end{figure}

To see this explicitly, consider the right-hand endpoint near $\theta = 0$. Expanding Eq.~\eqref{Eq:bubble_deformed} gives
\beq
x_R(\tau) = R_{\star} - \frac{R^2_{\star}}{3m_q}(a_t^2 + a_x^2) + i\frac{4R_{\star}}{3m_q}a_x a_t\, \tau + \cdots\, .
\eeq
After analytic continuation to Lorentzian time,
\beq
x_R(t) = R_{\star} - \frac{R^2_{\star}}{3m_q}(a_t^2 + a_x^2) - \frac{4R_{\star}}{3m_q}a_x a_t\, t + \cdots\, ,
\eeq
so the quark emerges with longitudinal velocity
\beq
v_0= -\frac{4R_{\star}}{3m_q} a_x a_t .
\label{Eq:v0}
\eeq
Similarly for the left-hand endpoint one finds
\beq
x_L(t)= -R_{\star} + \frac{R^2_{\star}}{3m_q}(a_t^2 + a_x^2) - \frac{4R_{\star}}{3m_q}a_x a_t\, t + \cdots\, .
\eeq
Both endpoints therefore acquire the same longitudinal velocity proportional to the mixed gradient term $a_x a_t$. The corresponding Lorentzian trajectories are shown in \Cref{Fig:bounce_deformation_Lorentzian}.

The deformation also modifies the initial quark-antiquark separation $L$. At nucleation,
\begin{align}
    L = x_{R}(0) - x_{L}(0) = 2R_{\star} - \frac{2R^2_{\star}}{3m_q}(a_t^2 + a_x^2).
\label{Eq:L_total}
\end{align}
Thus the separation is always smaller than the average diameter $2R_{\star}$ of the deformed bounce. Expressed instead relative to the undeformed ground-state radius $R_{0} = m_{q}/\kappa$,
\beq
L \simeq 2R_0 \left(1+\frac{2a_t^2-4a_x^2}{3\kappa}\right).
\eeq
Relative to the ground-state bounce, temporal gradients therefore increase the nucleation separation, whereas spatial gradients decrease it. Including a nonzero axion mass together with a homogeneous displacement $a = a_{\star}$ always acts to reduce $L$.

\subsection{Energy and momentum conservation}\label{sec:energyconservation}

It is instructive to examine how energy conservation is realised for the deformed bounce solution. For the undeformed critical bounce, the energy to produce the quark-antiquark pair comes from the segment of string that is removed. In the deformed case, the quarks cost energy
\begin{align}
    \Delta E_{\rm quarks} = 2m_q + m_q v_0^2 = 2m_q + \mathcal{O}(a_{x,t}^4),
\end{align}
and the missing string segment contributes
\beq
    \Delta E_{\rm gap} = -\kappa L = -2m_{q} - \frac{4m_{q}}{3 \kappa} (a_t^2-2a_x^2),
\eeq
which suggests an additional contribution at $\mathcal{O}(a_{x,t}^2)$. Indeed, the energy resides in the axion field, whose Lorentzian-signature worldsheet Hamiltonian density is
\beq
\mathcal H_{a} = \frac12 (\partial_{t} a)^2+\frac12 (\partial_{x} a)^2~.
\eeq
Before breaking, the string carries an energy density
\beq
{\mathcal H}^{(0)} = \frac12(a_t^2+a_x^2).
\eeq
At $t=0$, the Lorentzian continuation of the bounce solution gives
\begin{align}
\partial_{t} a^{(b)}(0,x)=a_t\left(1+\frac{R_{\star}^2}{x^2}\right),
\qquad
\partial_x a^{(b)}(0,x)=a_x\left(1-\frac{R_{\star}^2}{x^2}\right),
\end{align}
on the surviving string segments $|x|>R_{\star}$ (here we neglect higher-order corrections to the endpoint separation as these only contribute at yet higher order to the energy balance). The corresponding change in axion field energy between the original infinite string and the post-nucleation configuration is
\begin{align}
\Delta E_a = 2\int_{R_{\star}}^\infty \mathcal{H}^{(b)}\,dx - 2\int_0^\infty \mathcal{H}^{(0)}\,dx = \frac{4R_{\star}}{3} (a_t^2-2a_x^2) + \mathcal{O}(a_{x,t}^4).
\end{align}
In total, one finds
\beq
\Delta E_{\rm tot} = \Delta E_{\rm quarks} + \Delta E_{\rm gap} + \Delta E_{a} = 0 + \mathcal{O}(a_{x,t}^4).
\eeq
Energy conservation is therefore satisfied at leading nontrivial order, but only after including both effects of the axion background: the $\mathcal{O}(a_{x,t}^2)$ deformation of the bounce geometry, and the distortion of the axion profile through $\Delta_{a}$. Extending the calculation beyond this order would require incorporating the correction to $\Delta_{a}$ induced by the non-circular bounce profile. The leading-order treatment above is nevertheless sufficient for our purposes.

It is also worthwhile checking that momentum is conserved. At first glance, the common longitudinal velocity of the nucleated quarks seems to break momentum conservation,
\beq
\Delta P_{\rm quarks} = 2m_q v_0 \simeq -\frac{8}{3}R_{\star}\,a_x a_t + O(a_{x,t}^4).
\eeq
However, this is accounted for by the momentum density on the string when both temporal and spatial axion gradients are present. The initial unbroken string carries a Lorentzian-signature momentum density
\beq
{\mathcal P}^{(0)} = T^{01} = -(\partial_{t} a)(\partial_{x} a) = -a_t a_x\, ,
\eeq
while for the bounce solution, the momentum density on the remaining string segments at $t=0$ is
\beq
{\mathcal P}^{(b)}(x) = -(\partial_{t} a^{(b)}) (\partial_{x} a^{(b)}) = - a_t a_x\left(1-\frac{R_{\star}^4}{x^4}\right).
\eeq
The resulting change in the field momentum is then
\beq
\Delta P_{\rm field} = 2\int_{R_{\star}}^\infty \mathcal{P}^{(b)}\,dx - 2\int_0^\infty \mathcal{P}^{(0)}\,dx = \frac{8}{3}R_{\star}\,a_t a_x\, ,
\eeq
which precisely balances the momentum carried by the endpoint quarks. Momentum conservation is therefore also satisfied, provided one includes the momentum stored in the axion field configuration both before and after nucleation.

\section{Massive axion background}
\label{sec:generalsolution}

We now turn to the more general case $m_{a} \neq 0$. Unlike the massless theory, the resulting boundary-value problem is no longer sufficiently simple to admit a closed-form expression for $\Delta_{a}$ throughout the entire bounce geometry. Instead, we work perturbatively in the initial axion data and determine $\Delta_{a}$ only to leading order, and since the correction to the bounce action $\delta B_{a}$ is obtained from Eq.~\eqref{Eq:DeltaS_effE}, it is sufficient to determine $\Delta_{a}$ only in the asymptotic regions $|\tau| > R$.

We shall assume that the axion background varies on length scales much larger than the bounce size, which in particular implies the low-mass regime $m_{a} R_{\star} \ll 1$. In this regime the bounce only probes a small neighbourhood of the nucleation point, and the leading correction can depend only on the local value of the field and its first derivatives:
\begin{align}
    a^{(0)}(t,x) = a_{\star} + a_{x} x + a_{t} t + \cdots.
\end{align}
To compute the correction explicitly it is convenient to consider the representative solution
\begin{align}
    a^{(0)}(t,x) = [\alpha_{c} \cos(kx) + \alpha_{s}\sin(kx)][\beta_{c} \cos(\omega_{k} t) + \beta_{s}\sin(\omega_{k}t)], \quad \omega_{k}^{2} = m_{a}^{2} + k^{2},
\label{Eq:a0_solution}
\end{align}
which solves the axion equation of motion on the unbroken worldsheet. For this profile,
\begin{align}
    a_{\star} = \alpha_{c} \beta_{c}, \quad a_{x} = k\, \alpha_{s} \beta_{c},\quad a_{t} = \omega_{k} \alpha_{c} \beta_{s}.
\end{align}
The final result will be expressed entirely in terms of the local quantities $a_{\star}$, $a_{x}$, and $a_{t}$, and therefore applies to a general smooth background.

The corresponding Euclidean solution on the unbroken worldsheet is
\begin{align}
    a_{E}^{(0)}(\tau,x) = [\alpha_{c} \cos(kx) + \alpha_{s}\sin(kx)][\beta_{c}\cosh(\omega_{k}\tau) - i\beta_{s}\sinh(\omega_{k}\tau)].
\label{Eq:axion_E_nb}
\end{align}
The correction $\Delta_{a}$ on the bounce geometry satisfies Eqs.~\eqref{Eq:Delta_BC1}-\eqref{Eq:Delta_BC2}, together with the leading-order Neumann condition
\beq
\partial_r\Delta_{a}\big|_{r=R} =-\partial_r a_E^{(0)}\big|_{r=R}.
\label{Eq:Delta_NBC}
\eeq
These conditions define an exterior Neumann problem that may be solved using standard methods of potential theory. The solution can be written as a single-layer potential
\begin{align}
    \Delta_{a}(\tau,x) = \int_{r'=R}ds'\, G(\tau,x;\vec{x}') \sigma(\vec{x}'),
\label{Eq:Delta_solution}
\end{align}
where
\begin{align}
    G(\vec{x},\vec{x}') = \int \frac{dp}{2\pi} \frac{e^{- \omega_{p} | \tau - \tau'|}}{2 \omega_{p}} e^{i p (x-x')}
\end{align}
is the Green's function for $\nabla^{2} - m_{a}^{2}$ that satisfies the same boundary conditions as $\Delta_{a}$ at $\tau = \pm T/2$ (provided $|\tau'| < T/2$)\footnote{This restriction is to be expected. Since $\Delta_{a}$ is a single-layer potential supported on the circle $r=R$, the standard jump relations imply that its normal derivative is discontinuous across this boundary. Hence, the solution is non-analytic at $r=R$, and contour deformations cannot be continued through this point without accounting for the associated jump contribution (equivalently interpretable as a contact term with the source density $\sigma$). The complex time contour is therefore restricted to the region $T/2 > R$.}. In fact, this is nothing more than the modified Bessel function
\begin{align}
    G(\vec{x},\vec{x}') = \frac{1}{2\pi}K_{0}\left(m_{a} \sqrt{(\tau - \tau')^{2} + (x-x')^{2}} \right).
\end{align}
The source density $\sigma(\vec{x}')$ is fixed by the Neumann condition Eq.~\eqref{Eq:Delta_NBC}. Using the Bessel addition formula
\begin{align}
    K_{0}\left(m_{a} \sqrt{(\tau - \tau')^{2} + (x-x')^{2}} \right) = \sum_{n\in \mathbb{Z}} I_{|n|}(m_{a}r') K_{|n|}(m_{a}r)e^{in(\theta - \theta')},
\end{align}
valid when $r>r'$, and defining the Fourier components
\begin{align}
    \sigma_{n} = \frac{1}{2\pi} \int_{0}^{2\pi} d\theta' \sigma(\theta') e^{in\theta'}, \qquad \Delta_{a,n}(r) = \frac{1}{2\pi} \int_{0}^{2\pi} d\theta\, \Delta_{a}(r,\theta) e^{in\theta},
\end{align}
the solution Eq.~\eqref{Eq:Delta_solution} reduces to
\begin{align}
    \Delta_{a,n}(r) = R\,I_{|n|}(m_{a}R) K_{|n|}(m_{a}r) \sigma_{n},
\end{align}
so that
\begin{align}
    \sigma_{n} = \frac{\partial_{r} \Delta_{a,n}(R)}{m_{a}R\, I_{|n|}(m_{a}R) K_{|n|}'(m_{a}R)}\, .
\end{align}
We now specialise to the long-wavelength regime $\omega_{k} R \ll 1$. Expanding the denominator
\begin{align}
    m_{a}R\, I_{|n|}(m_{a}R) K_{|n|}'(m_{a}R) = \begin{cases}
        -1 \ \   + \mathcal{O}((m_{a}R)^{2}\ln(m_{a}R)) & n=0\\
        -1/2 + \mathcal{O}((m_{a}R)^{2}\ln(m_{a}R)) & |n|=1\\
        -1/2 + \mathcal{O}((m_{a}R)^{2}) & |n| \geq 2.
    \end{cases}
\end{align}
These constant terms are sufficient to obtain the leading correction to the bounce action. The numerator follows from Eq.~\eqref{Eq:Delta_NBC} together with the known background solution Eq.~\eqref{Eq:axion_E_nb}. Expanding in $\omega_{k} R$ yields
\begin{align}
    \partial_{r} \Delta_{a,0} &= -\frac{1}{2} m_{a}^{2} a_{\star}R + \mathcal{O}((m_{a}R)^3), \nonumber\\
    \partial_{r} \Delta_{a,1} &= -\frac{1}{2} (a_{x} + a_{t}) + \mathcal{O}((\omega_{k}R)^2), \\
    \partial_{r} \Delta_{a,-1} &= -\frac{1}{2} (a_{x} - a_{t}) + \mathcal{O}((\omega_{k}R)^2), \nonumber
\end{align}
while all higher harmonics are suppressed by at least one additional power of $\omega_{k}R$. Retaining only the leading contributions therefore gives
\begin{align}
    \sigma(\theta') \simeq \frac{1}{2} m_{a}^{2} a_{\star}R + 2a_{x} \cos \theta' - 2i a_{t} \sin \theta'.
\end{align}
Substituting this result into
\begin{align}
    \Delta_{a}(\tau,x) = \int_{0}^{2\pi}d\theta'\, R \int \frac{dp}{2\pi} \frac{e^{- \omega_{p} | \tau - R \sin \theta'|}}{2 \omega_{p}} e^{i p (x-R \cos \theta')} \sigma(\theta')
\end{align}
and expanding to leading order in $\omega_{k}R$ gives
\begin{align}
    \Delta_{a}(\tau,x) \simeq \begin{cases}\pi R^{2} \int \frac{dp}{2\pi} \frac{e^{-\omega_{p} \tau} e^{ipx}}{\omega_{p}} (\frac{1}{2}m_{a}^2 a_{\star} - i\omega_{p} a_{t} - ipa_{x}) \quad \text{for}\quad \tau > +R,\\
    \pi R^{2} \int \frac{dp}{2\pi} \frac{e^{+\omega_{p} \tau} e^{ipx}}{\omega_{p}} (\frac{1}{2}m_{a}^2 a_{\star} + i \omega_{p} a_{t} - ipa_{x}) \quad \text{for}\quad \tau < -R.
    \end{cases}
\label{Eq:Delta_approx}
\end{align}
This expression is valid when integrated against $a_{E}^{(0)}$, since this projects onto $p = \pm k$ and we have assumed $\omega_{k} R \ll 1$.

Finally, substituting Eq.~\eqref{Eq:Delta_approx} into Eq.~\eqref{Eq:DeltaS_effE} yields the axion contribution to the bounce action
\begin{align}
    \delta B_{a} = - \pi R^{2} \left[a_{x}^{2}- a_{t}^{2} + \frac{1}{2} m_{a}^{2} a_{\star}^{2}\right] + \mathcal{O}((\omega_{k} R)^{3}).
\end{align}
This confirms that in the long-wavelength regime, the leading correction depends only on the local value of the axion field and its first derivatives at the nucleation point. Comparing with the functional form obtained in the massless case, Eq.~\eqref{Eq:B_a_massless}, we identify the effective tension
\begin{align}
\kappa_{\rm eff} =  \kappa  + a_{x}^{2}- a_{t}^{2} + \frac{1}{2} m_{a}^{2} a_{\star}^{2}.
\end{align}
This confirms the more general expression quoted in Eq.~\eqref{eq:effective_tension}, and it follows that the critical radius, bounce action, and decay rate depend on the axion only through this effective tension as found in \Cref{sec:simplecases}.

The calculation also clarifies the origin of the relative factor of two between the mass and derivative terms appearing in $\kappa_{\rm eff}$: the mass contribution arises from the monopole component of the source density $\sigma(\theta')$, whereas the derivative contributions originate from its dipole moments. Finally, as in the massless case, this result is independent of the contour parameter $T$, providing a non-trivial consistency check of the calculation.

The analysis presented here assumes that the axion varies over length scales much larger than the bounce radius $R_{\star}$, which requires in particular $m_{a} R_{\star} \ll 1$. At the same time, the semi-classical expansion requires a large bounce action $B_{0} = \pi m_{q}^{2} / \kappa \gg 1$. Using the lattice value of the axion mass $m_{a} \simeq 1.85 \sqrt{\kappa}$, one finds $m_{a} R_{\star} \simeq 1.85\, m_{q} /\sqrt{\kappa} \gg 1$ for small deformations $R_{\star} \simeq m_{q}/\kappa$. Hence, realistic QCD strings lie outside the strict domain of validity of this analysis. In the more realistic regime $m_{a} R_{\star} \sim \mathcal{O}(1)$, the bounce shape will differ substantially from the small-deformation solution, and neither the zeroth-order bubble radius nor the corresponding decay exponent can be quantitatively trusted. Extending the analysis to the regime $m_{a} R_{\star} \not\ll 1$ requires going beyond the long-wavelength expansion developed here. We are currently investigating this problem using a combination of analytical and numerical techniques, and will report the result, together with other extensions of the present work, elsewhere.

\section{Conclusion}
\label{sec:Conclusions}

We have computed the leading effect of a semi-classical worldsheet pseudoscalar (``axion") field excitation on the string-breaking rate of a metastable string. Our analysis was based on Euclidean path integral methods formulated on a Schwinger-Keldysh complex-time contour. To our knowledge, this is the first calculation of string breaking in the presence of a non-trivial excited worldsheet background beyond the Nambu-Goldstone sector carried out within a theoretically controlled framework.

The principal result is that the axion modifies the decay rate exponent Eq.~\ref{Eq:Lund} through the replacement of the string tension by an effective tension, Eq.~\ref{eq:effective_tension}. Notably, this effective tension is not simply the local energy density on the string. Rather, it is a worldsheet Lorentz scalar constructed from the axion field and its first derivatives, together with the field amplitude in the massive case. Spatial gradients and large  amplitudes exponentially enhance the breaking rate, whereas temporal gradients exponentially suppress it. By contrast, Nambu-Goldstone excitations in the corresponding long-wavelength regime affect only the prefactor of the decay rate~\cite{Monin:2008mp}. The axion contribution therefore constitutes the leading modification to string breaking arising from worldsheet excitations in this regime.

Our analysis also reveals several interesting features of the decay process. For a string in its ground state, decay proceeds through the Euclidean bounce geometry consisting of a circular hole excised from the worldsheet. Upon analytic continuation, this configuration describes the nucleation of a quark-antiquark pair at rest, followed by uniform acceleration under the string tension. In an excited axion background, the bounce is deformed away from circularity, leading to a modified nucleation separation. The Euclidean bounce generally becomes a complex-valued curve, while still continuing to a real Lorentzian solution. The imaginary part of the Euclidean geometry induces equal non-zero initial velocities for the nucleated quark endpoints. Energy and momentum are seen to be conserved in the decay process once the modification of the axion profile on the surviving string segments after nucleation is properly taken into account, the effect of the bounce on the axion field extending outside of the quark-antiquark boundary curve.

As discussed in \Cref{sec:generalsolution}, phenomenologically relevant applications require extending the present analysis beyond the small-axion-mass regime, $m_{a} R_{\star} \ll 1$, where $R_{\star}$ is the naive leading-order critical bubble radius. The methods developed in \Cref{Sec:Keldysh,Sec:bounceexponent} provide a framework for doing so through a combination of analytic and numerical techniques, and we are investigating this regime with results to be presented elsewhere.

Despite this limitation, the results obtained here already point to potentially important consequences for QCD phenomenology. Because the decay rate exponent depends sensitively on both the effective tension and on the mass of the nucleated quarks, worldsheet excitations can modify flavour production during hadronisation, including relative rates of strange- and light-quark production. We explore these effects in a companion paper~\cite{StringsII}, where we also address the statistical properties of the excitation spectrum of confining flux tubes produced in high-energy collisions.

As one further application, the formalism developed here enables a first-principles calculation of the transverse momentum distribution of the produced quarks in the presence of both worldsheet Nambu-Goldstones and the axion. This topic is developed in a second companion work~\cite{TransverseMomentum}.

Our results demonstrate that excited states of the confining string can significantly modify string breaking behaviour. Incorporating such effects may therefore be essential for a quantitatively accurate and systematically improvable description of hadronisation based on the notable success of the Lund string model~\cite{Andersson:1983ia,Sjostrand:2014zea}. At the same time, the methods developed here are considerably more general and may be applied to tunnelling and phase-transition processes in excited backgrounds across a wide range of physical systems.

\section{Acknowledgements}

We thank Mike Teper for valuable discussions on the effective theory of confining flux tubes.  We especially wish to thank Jade Abidi, Javira Altmann, Nat{\'a}lie Koscelansk{\'a} van IJcken, and Peter Skands for their ongoing collaboration and fruitful exchange of ideas. EC is supported by the Clarendon Fund Scholarship in partnership with the Oxford-Berman Graduate Scholarship, and thanks Sergey Sibiryakov for valuable discussions. JMR thanks Sergei Dubovsky for previous discussions of the effective string theory of the confining flux tube, and the theory group at NYU for their hospitality.

\appendix

\bibliographystyle{JHEP}
\bibliography{Bibliography_first}

\end{document}